\documentclass[twocolumn,amsmath,amssymb,aps,prb,floatfix,superscriptaddress]{revtex4-2}
\usepackage[latin1,utf8]{inputenc}
\usepackage{epsfig}
\usepackage{graphicx}
\usepackage{multirow}
\usepackage{color}
\usepackage{latexsym}
\usepackage[Large,FIGTOPCAP]{subfigure}
\usepackage{IEEEtrantools}
\usepackage{blkarray}
\usepackage{mathtools}
\usepackage[colorlinks=true,allcolors=blue]{hyperref}
\begin{document}

\title{A model for coupled $4f-3d$ magnetic spectra: a neutron scattering study of the Yb--Fe hybridisation in Yb$_3$Fe$_5$O$_{12}$}

\author{Viviane~Pe\c canha-Antonio}
\email{viviane.pecanhaantonio@physics.ox.ac.uk}
\author{Dharmalingam Prabhakaran}
\affiliation{Department of Physics, University of Oxford, Clarendon Laboratory, Oxford OX1 3PU, United Kingdom}
\author{Christian~Balz}
\author{Aleksandra~Krajewska}
\affiliation{ISIS Neutron and Muon Source, STFC Rutherford Appleton Laboratory, Didcot OX11 0QX, United Kingdom}
\author{Andrew~T.~Boothroyd}
\email{andrew.boothroyd@physics.ox.ac.uk}
\affiliation{Department of Physics, University of Oxford, Clarendon Laboratory, Oxford OX1 3PU, United Kingdom}

\date{\today}

\begin{abstract}
In this work, we explore experimentally and theoretically the spectrum of magnetic excitations of the Fe$^{3+}$ and Yb$^{3+}$ ions in ytterbium iron garnet (Yb$_3$Fe$_5$O$_{12}$). We present a complete description of the crystal-field splitting of the $4f$ states of Yb$^{3+}$, including the effect of the exchange field generated by the magnetically ordered Fe subsystem. We also consider a further effect of the Fe--Yb exchange interaction, which is to hybridise the Yb crystal field excitations with the Fe spin-wave modes at positions in the Brillouin zone where the two types of excitations cross. We present detailed measurements of these hybridised excitations, and propose a framework which can be used in the quantitative analysis of the coupled spectra in terms of the anisotropic $4f-3d$ exchange interaction.
\end{abstract}

\maketitle

\section{Introduction}

Iron garnets are a family of compounds with chemical formula RE$_3$Fe$_5$O$_{12}$, where RE represents a trivalent rare-earth or yttrium (Y) ion. Y$_3$Fe$_5$O$_{12}$ (YIG) is the most prominent member of this family, having facilitated the recent development of research fields such as spintronics \cite{PhysRevApplied.4.047001}, magnonics \cite{Chumak}, and hybrid quantum information systems \cite{Nakamura}. Iron garnets incorporating magnetic rare-earth ions have also been studied for many years for their interesting spin transport phenomena \cite{Geprags}, magnetoelectric properties \cite{Hur}, magneto-optical effects \cite{Yoshimoto} and anisotropic magnetization which varies according to the choice of RE ion \cite{Pearson}.

\begin{figure}
\begin{tabular}{ll}
\begin{minipage}{0.47\textwidth}
\hspace{-2.em}
\subfigure[]{%
\includegraphics[trim=200 0 200 0, clip,width=0.60\textwidth]{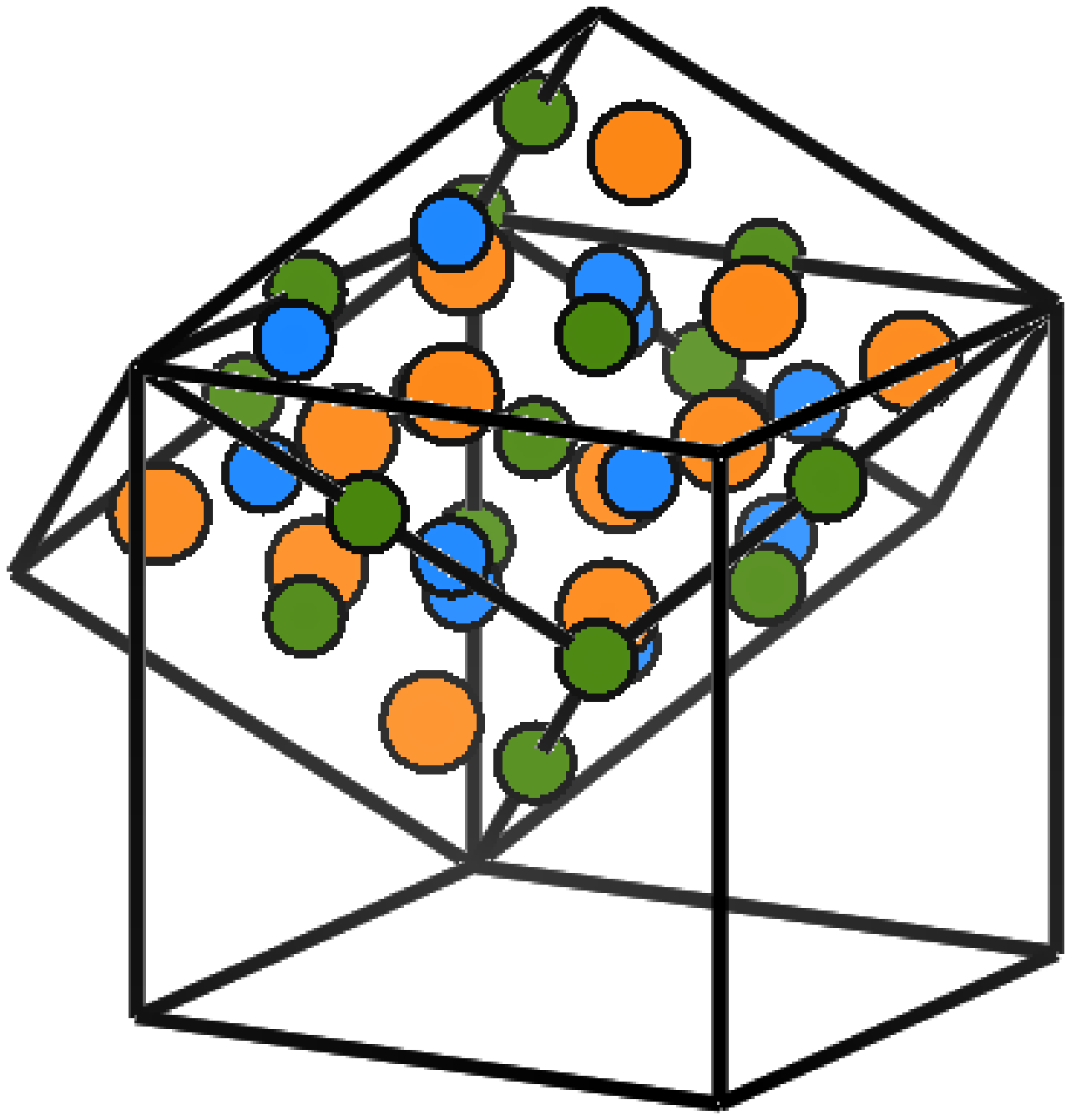}\label{fig1-a}}
\hspace{-1.5em}
\subfigure[]{%
\includegraphics[trim=300 50 300 50, clip,width=0.46\textwidth]{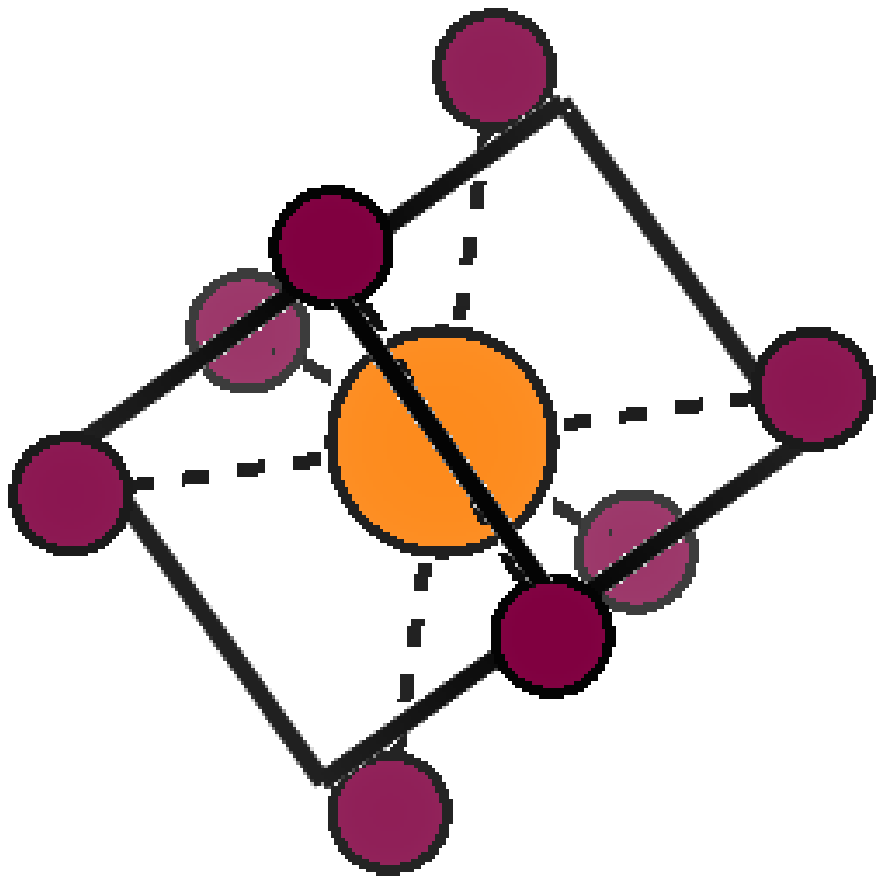}\label{fig1-c}}\\
\subfigure{%
\includegraphics[trim=0 0 0 0, clip,width=1.0\textwidth]{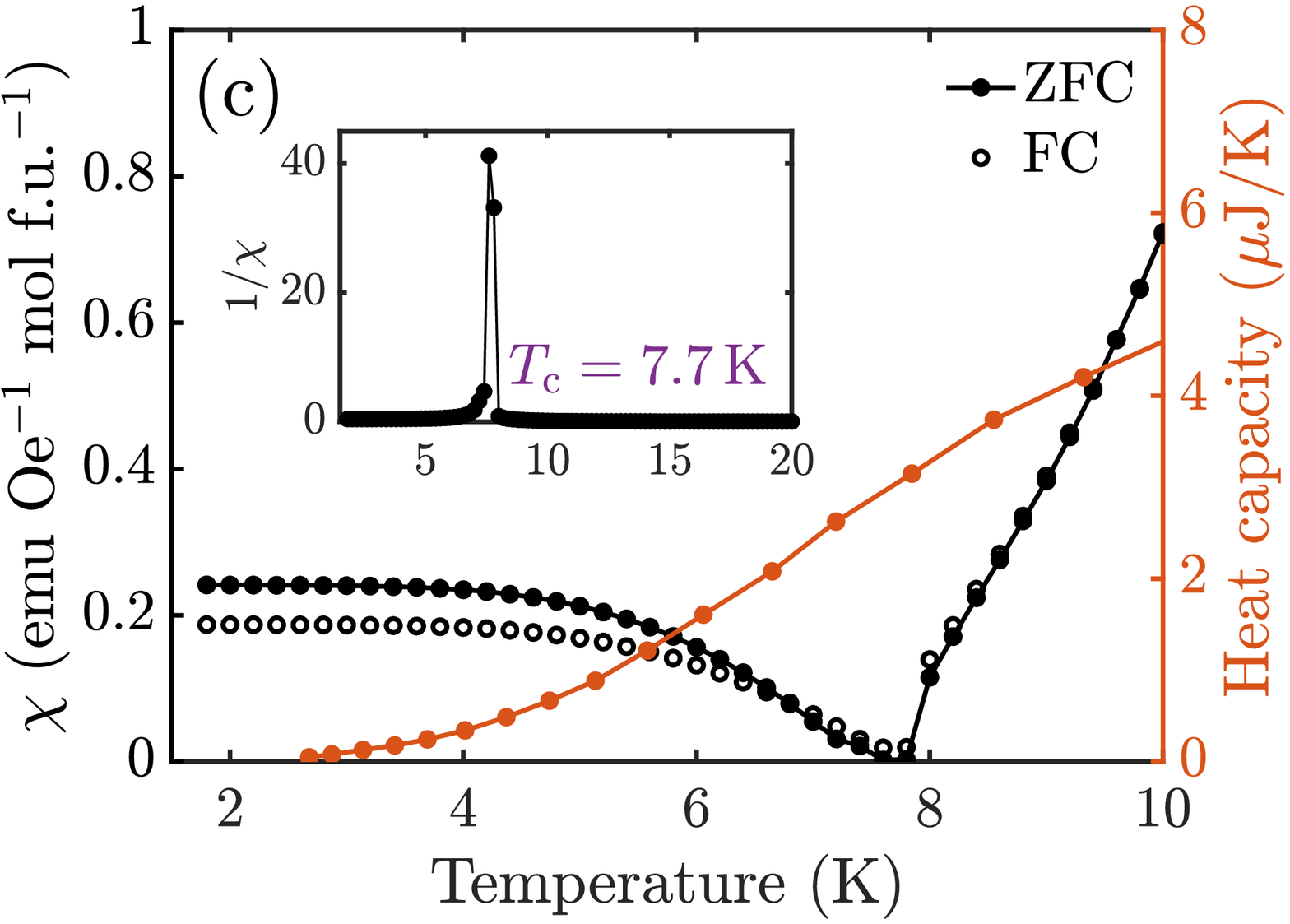}\label{fig1-b}}
\end{minipage}
\end{tabular}
\caption{\subref{fig1-a} Garnet \emph{primitive} unit cell, displayed within the conventional cubic vertices. Trivalent rare-earths are shown in orange, while green and blue circles correspond to Fe$^{3+}$ at $24d$ and $16a$ positions, respectively. Oxygens are omitted in \subref{fig1-a} for clarity. \subref{fig1-c} Pseudocubic oxygen environment of the Yb$^{3+}$, as seen along one of the equivalent cubic $\langle 0 0 1 \rangle$ directions. The point symmetry of the central atom is $222$. \subref{fig1-b} Magnetic susceptibility $\chi$ for Yb$_3$Fe$_5$O$_{12}$, measured in a magnetic field of 100\,Oe applied along the $[110]$ direction. The compensation temperature $T_\textup{c}$ occurs at about 7.7\,K. Heat capacity data are plotted along with $\chi$ in order to emphasise that the compensation temperature is not accompanied by a phase transition. The inset in \subref{fig1-b} shows the inverse susceptibility over the same temperature interval, where the crossover at $T_\textup{c}$ can be more clearly seen.}
\label{fig1}
\end{figure}

Single crystals of YIG can be grown with pristine quality, a characteristic which has contributed to the compound's importance in basic research \cite{Cherepanov}. The iron garnets crystallise in a body-centred cubic structure described by the space group $Ia\bar{3}d$ (no.~230). The primitive unit cell is remarkably full, containing a total of four RE$_3$Fe$_5$O$_{12}$ formula units. The RE$^{3+}$ ions occupy the $24c$ Wyckoff positions, while Fe$^{3+}$ ions are distributed over the two symmetry-inequivalent $16a$ and $24d$ sites [see Fig.~\ref{fig1}\subref{fig1-a}].

YIG is ordered magnetically below $T_\textup{N}=560$\,K, with a $\mathbf{k}=0$ structure and Fe$^{3+}$ spins ($S=5/2$) aligned along one of the cubic $\langle 111 \rangle$ directions \cite{Cherepanov}. Spins located at equivalent crystallographic positions are ordered parallel to one other, while spins in the $16a$ sites point in the opposite direction to those in the $24d$ sites. The unequal number of magnetic moments aligned in opposite directions (8 against 12 in one primitive unit cell) creates the net magnetisation responsible for textbook ferrimagnetism. A manifestation of the coherent precession of the coupled Fe magnetic moments, spin waves in YIG are important to the material's exceptional spin transport properties. As there are 20 Fe ions in the primitive unit cell, there are 20 spin-wave modes for each reciprocal lattice wavevector. The spin-wave spectrum extends in energy up to about 85\,meV. \cite{Princep,PhysRevB.97.054429}. 

In recent years, renewed interest in the magnon degrees of freedom of YIG followed the experimental observation of the \emph{Spin Seebeck effect} (SSE) \cite{Uchida1,Uchida2,PhysRevX.4.041023}. The SSE presents an interesting, energetically economic way of harnessing the magnetic excitations in an insulator in order to produce an electric current in a metal attached to it \cite{PhysRevApplied.4.047001}. Several characteristics of Y$_3$Fe$_5$O$_{12}$ demonstrate how appropriate the material seems to be for applications based on the spin Seebeck effect, particularly its extremely low (in fact, the lowest known \cite{Cherepanov}) spin-wave damping, responsible for pure magnon diffusion over distances up to $\sim10\ \mu\textup{m}$ at room temperatures \cite{PhysRevB.92.224415}.

It has been shown that one of these spin-wave branches, an acoustic mode isolated in energy up to about $T\sim300$\,K ($E=k_\textup{B}T\sim26$\, meV)\footnote{This branch is somewhat confusingly called the \emph{ferromagnetic} mode in the literature \cite{PhysRevB.89.024409}, even though the main interactions in YIG up to the fourth nearest-neighbours are antiferromagnetic \cite{Cherepanov,Princep,PhysRevB.97.054429,PhysRevB.92.054436}}, bears fundamental importance on the physics governing the SSE \cite{PhysRevLett.117.217201}. One way to microscopically manipulate the interactions responsible for the propagation of this mode in YIG is to substitute Y by a magnetic trivalent rare-earth ion. In this particular case, not only Fe--Fe interactions, but also rare-earth single-ion excitations, as well as RE--Fe coupling, will all contribute to the garnet magnetic spectrum in an energy range relevant to potential applications based on the SSE. 

Most of the discussion about YIG is also applicable to the RE iron garnets in general. Magnetic properties of the rare-earth substituted compounds were described shortly after their discovery in the late 1950s \cite{Geller1957}. Although the N\'eel temperature marking the long-range order of the Fe spins remains close to 560\,K for the whole series, iron garnets for which the RE sites are occupied by lanthanides heavier than Gd$^{3+}$ display a susceptibility compensation temperature $T_{\textrm{c}}$, when the bulk magnetisation of the compound vanishes \cite{PhysRev.137.A1034}. This point is understood to mark the temperature at which the ordered magnetic moment of the RE sublattice is identical to that of the Fe sublattices, but with opposite relative orientation. Below $T_{\textrm{c}}$, for certain RE, the magnetic arrangement of the RE atoms displays a canting around the $\langle 111 \rangle$ direction, forming a so-called umbrella structure \cite{Guillot1984,Hock1990,Hock1991,Tcheou}.

The aim of this work is to investigate how RE-Fe interactions modify the already largely understood spin-wave spectrum of YIG. We make use of inelastic neutron scattering to map in detail the reciprocal space of a single crystal of Yb$_3$Fe$_5$O$_{12}$ (YbIG). For this compound, $T_\textrm{c} \simeq 7.7$\,K, as the magnetic susceptibility data shown in Fig.~\ref{fig1}\subref{fig1-b} demonstrate. The Yb$^{3+}$ ($4f^{13}$) ion is chosen for the relative simplicity of its electronic shell when compared with other lanthanides. With one electron vacancy in the $f$ orbital, the Yb$^{3+}$ has orbital and spin quantum numbers $L=3$ and $S=1/2$. The spin-orbit ground state $J=L+S=7/2$, in the presence of a time reversal symmetry breaking potential, can thus be split into a maximum of eight levels.

In addition to developing a microscopic description of the $4f-3d$ hybridisation and low energy magnetic degrees of freedom in YbIG, we are also motivated by the desire to move a step forward in the modelling of the spin excitation spectra measured by inelastic neutron scattering. While linear spin-wave theory is a very well established method for treating coupling between effective spins-$\tfrac{1}{2}$, there have been relatively few attempts to model the hybridisation of single-ion and collective excitations involving several coupled orbital and spin states. The experiment performed in this work has demonstrated that such an understanding is necessary to fully explain the exchange interactions between $d$ and $f$ electronic configurations in Yb$_3$Fe$_5$O$_{12}$. This work is made possible by the exceptional resolution of inelastic neutron scattering at low energies ($<5$\,meV), which allows the observation, and subsequent description, of the $d-f$ exchange coupling in unprecedented detail.

\section{Overview of the magnetic spectrum of Ytterbium iron garnet}

The dominant term in the Hamiltonian of rare-earth insulators is usually the spin--orbit coupling $\mathcal{H}^\textup{SO} =\lambda \mathbf{L} \cdot \mathbf{S}$, where $\lambda$ is the spin-orbit parameter and $\mathbf{L}, \mathbf{S}$ are the orbital and spin angular momentum operators. For the Kramers ion Yb$^{3+}$, the spin--orbit coupling splits the $4f^{13}$ electronic configuration into two levels with total angular momentum quantum numbers $J=7/2$ and $5/2$, respectively, separated in energy by $7\lambda/2 \sim 1300$\,meV.

The next most relevant contribution is the crystalline electric field (CF) potential $\mathcal{H}^\textup{CF}$, which is responsible for splitting each $J$ multiplet of the Kramers Yb$^{3+}$ into $J+1/2$ doublets. The CF Hamiltonian is totally determined by the local symmetry of the rare-earth in the lattice. Each Yb in the garnet is coordinated by eight O$^{2-}$ anions, forming an environment which can be mapped onto a distorted cube [see Fig.~\ref{fig1}\subref{fig1-c}]. The point symmetry of the rare-earth in this arrangement is orthorhombic (point group $222$). The energy scale of the CF splitting of the $J=7/2$ multiplet in YbIG is about 80\,meV, as will be shown later.

The rare-earth is additionally subject to an exchange field from the ordered Fe-spins, which tends to align the Yb moments in a direction opposite to the net magnetisation of the Fe-sublattices. The effects of this local effective field upon the Yb$^{3+}$ single-ion levels are two-fold. First, the exchange field breaks the time-reversal symmetry (and, therefore, the Kramers degeneracy) of the Yb$^{3+}$ doublets. Second, it also breaks the orthorhombic $222$ point-group symmetry of the pure crystalline electric field potential. The latter occurs because, even though all the Yb ions experience the same CF, the orientation of the local CF-axes differs among the Yb sites relative to the direction of the Fe-exchange field. Consequently, the twelve rare-earth sites in the primitive unit cell are divided into two groups of six symmetry-equivalent positions \cite{PhysRev.124.1401}. These two groups display slightly different sets of Yb single-ion levels.

\begin{figure}
\centering
\includegraphics[trim=35 0 20 0, clip,width=8.2cm]{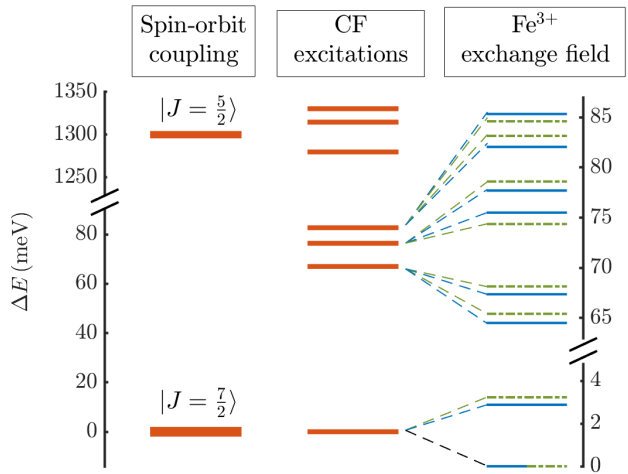}
\caption{Evolution of the single-ion Yb$^{3+}$ spectrum in YbIG. In the absence of time-reversal symmetry breaking, all the CF levels are Kramers doublets. The exchange field caused by the magnetic ordering of the Fe$^{3+}$ sublattice, as much as breaking the Kramers degeneracy, also splits the twelve rare-earth sites into two groups. On the right-hand side panel, the ground state multiplet ($|J=7/2\rangle$) excitations for each of those two sites (continuous and dashed lines) are shown.}\label{fig2}
\end{figure}

In summary, the ground state $J=7/2$ manifold of each of the two groups is split into $(2J+1)=8$ singlets. Therefore $2\times 8 = 16$ levels, including 14 excited states, may be identified in the spectrum. An overview of the single-ion spectrum of the Yb$^{3+}$ in the iron garnet structure may be found in Fig.~\ref{fig2}. 

\section{Experimental details}

Single crystals of Yb$_3$Fe$_5$O$_{12}$ were grown by the floating-zone method. X-ray and neutron Laue diffraction were used to select and orient crystals of high crystalline quality. Susceptibility and heat capacity measurements were performed on a Quantum Design MPMS3 magnetometer and on a Physical Property Measurement System (PPMS). 

Time-of-flight neutron-scattering experiments were performed on the MAPS spectrometer \cite{MAPS} at the ISIS facility. Three single crystals with a total mass of 3\,g were co-aligned with a resulting mosaic spread of less than $2^\circ$. The sample was mounted in a closed-cycle refrigerator (CCR), and data were recorded at the cryostat base temperature (approximately 6\,K). The sample was rotated around the cubic $[1 \bar{1} 0 ]$ axis in $1^\circ$ steps. The instrument chopper was operated in repetition-rate multiplication (RRM) mode and angular scans with incident energies $E_\textrm{i}$ of 25 and 120\,meV were performed simultaneously. The full width at half maximum (FWHM) of the energy resolution of these measurements is approximately 4\% of $E_\textrm{i}$ at zero energy transfer, decreasing to $\sim2\%$ for energy transfers close to 100\,meV.

Low energy data were collected on the time-of-flight spectrometer LET \cite{LET}, also at the ISIS facility. One single crystal of mass 1\,g selected from the MAPS sample was fixed in an aluminium mount and loaded in a helium cryostat. During data collection, at a temperature of 1.8\,K, the sample was rotated through an angle of $140^\circ$ in $1^\circ$ steps around the cubic $[1 \bar{1} 0 ]$ axis. RRM enabled the simultaneous measurement of $E_\textrm{i}=5.5$ and 17.3\,meV, (the latter with less flux) among other incident energies. The FWHM of the energy resolution at the elastic line is 0.16\,meV, down to about 0.05\,meV at 5\,meV energy transfer. 

\section{Experimental results}

\begin{figure}[h]
\centering
\subfigure{
\includegraphics[trim=0 0 0 0, clip,width=4.2cm]{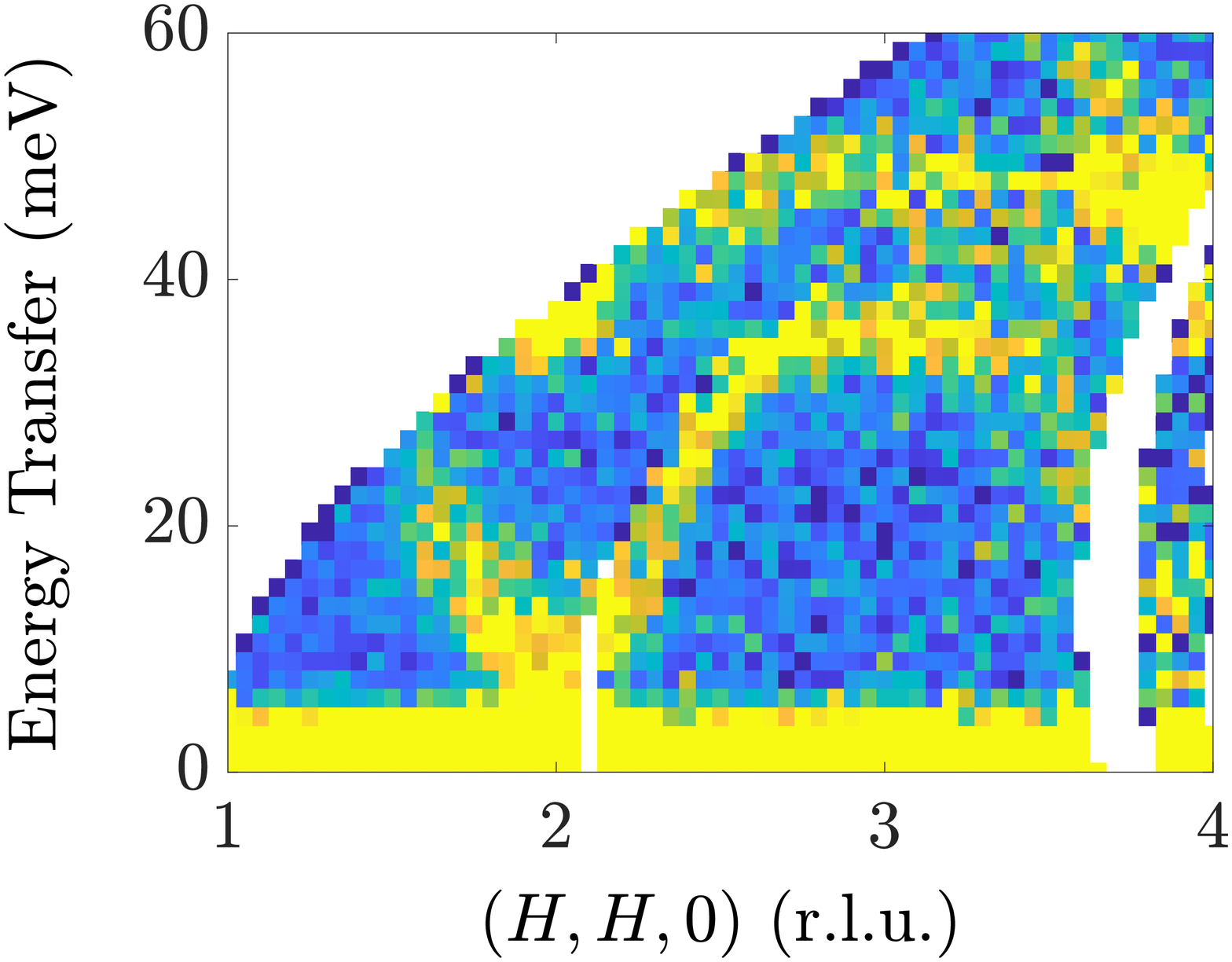}\label{fig3-e}}\hspace{-0.5em}
\subfigure{
\includegraphics[trim=0 0 0 0, clip,width=4.2cm]{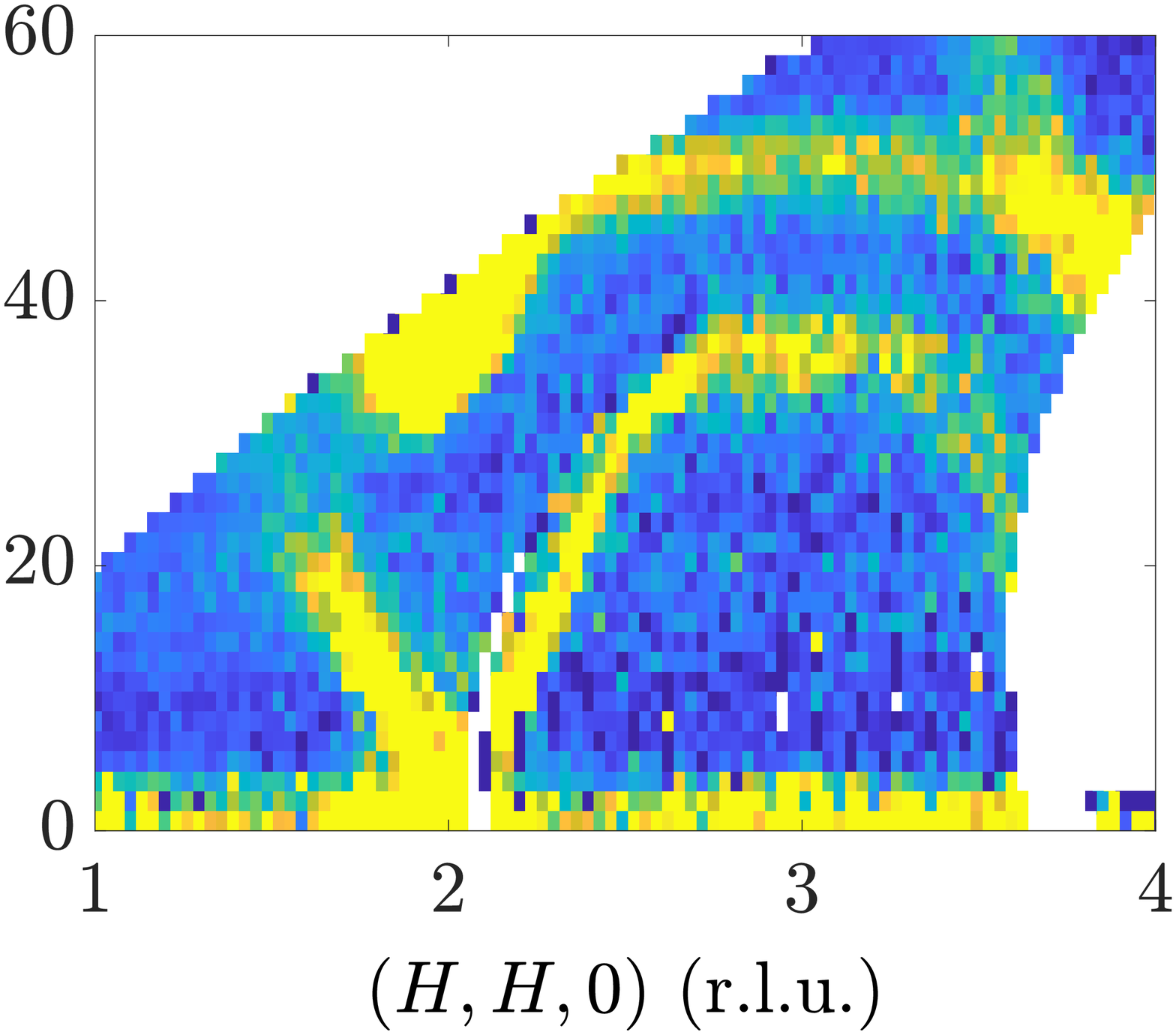}\label{fig3-f}}\vspace{-0.5em}
\subfigure{
\includegraphics[trim=0 0 0 0, clip,width=4.2cm]{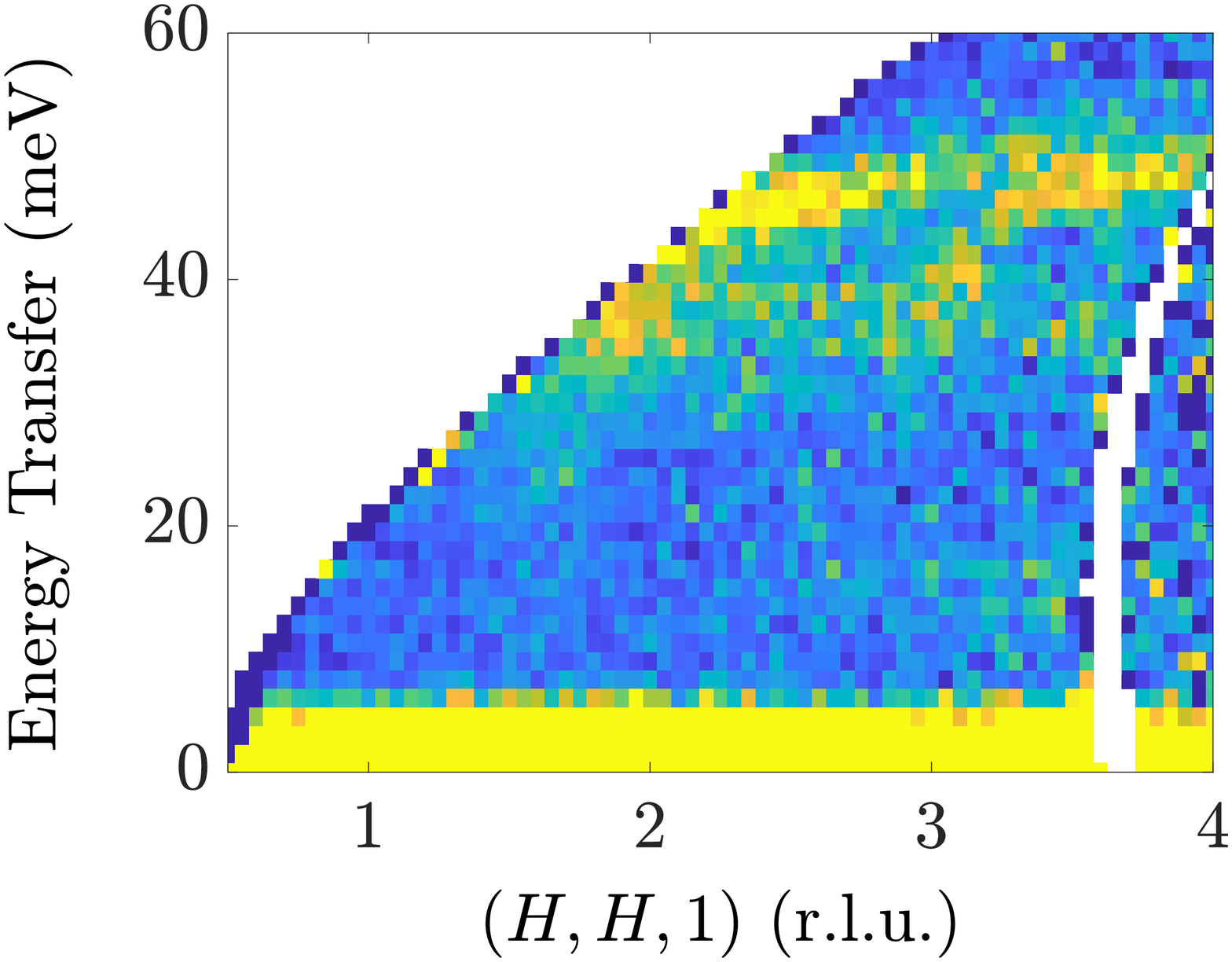}\label{fig3-g}}\hspace{-0.5em}
\subfigure{
\includegraphics[trim=0 0 0 0, clip,width=4.2cm]{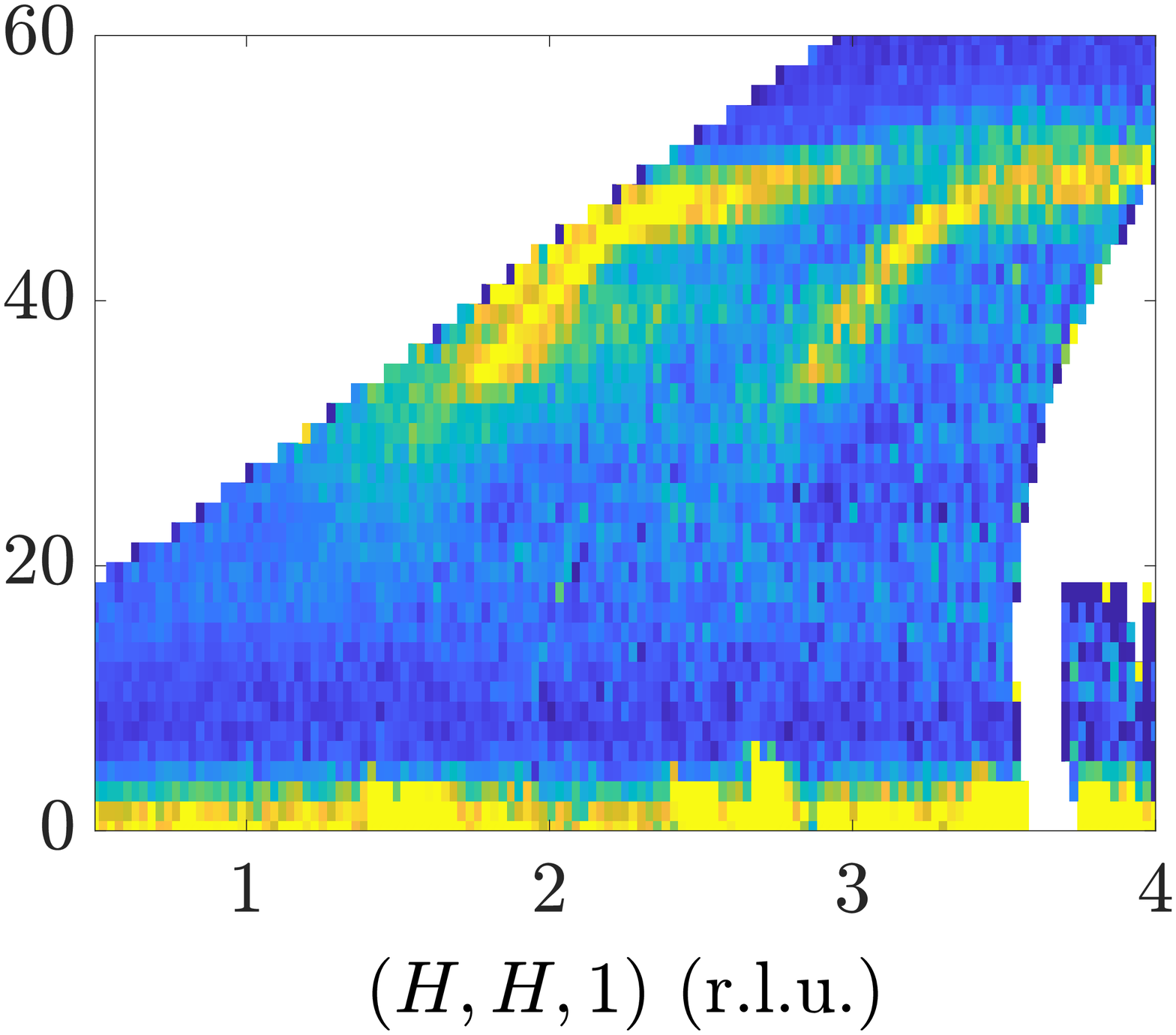}\label{fig3-h}}\vspace{-0.5em}
\subfigure{
\includegraphics[trim=0 0 0 0, clip,width=4.2cm]{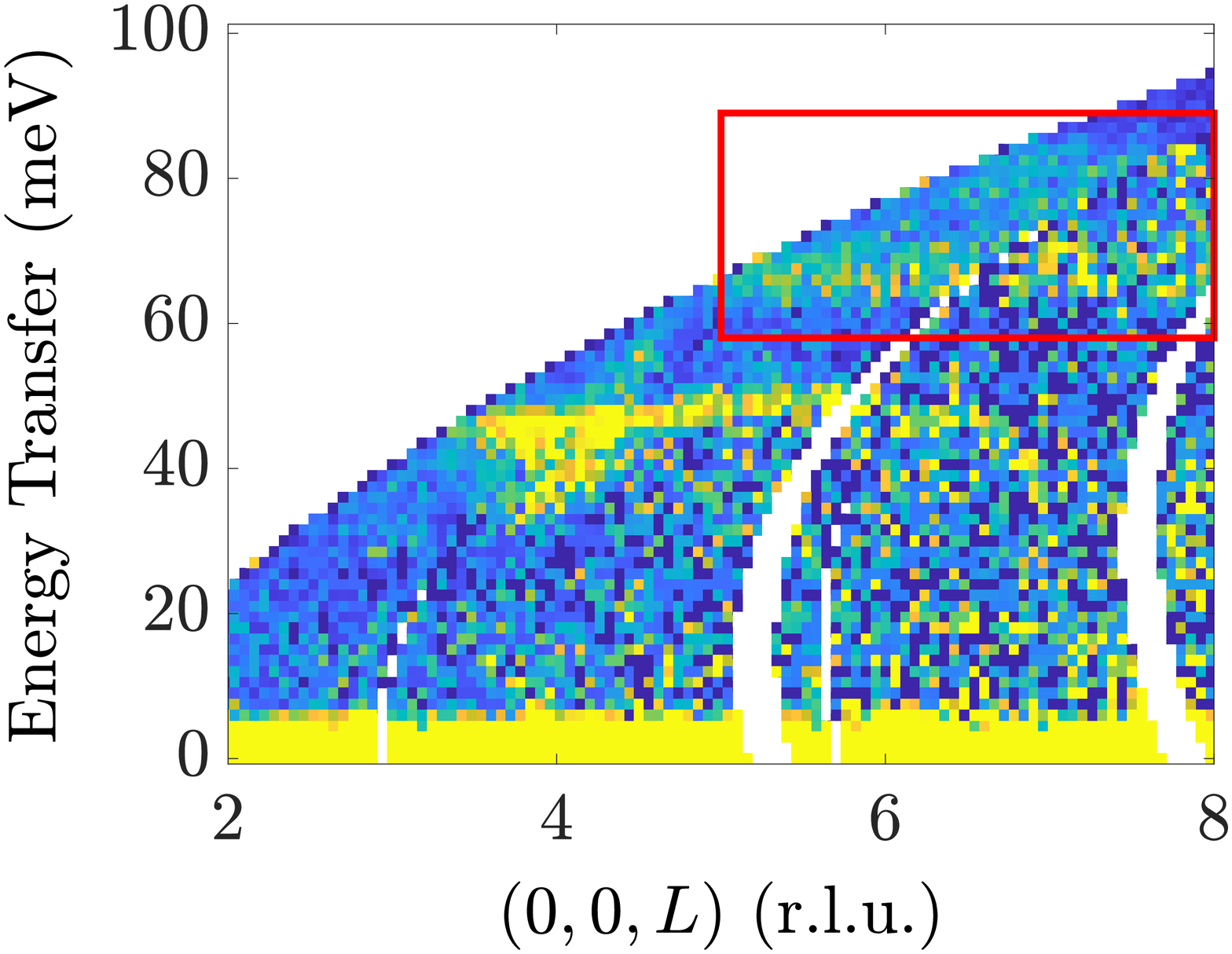}\label{fig3-a}}\hspace{-0.5em}
\subfigure{
\includegraphics[trim=0 0 0 0, clip,width=4.2cm]{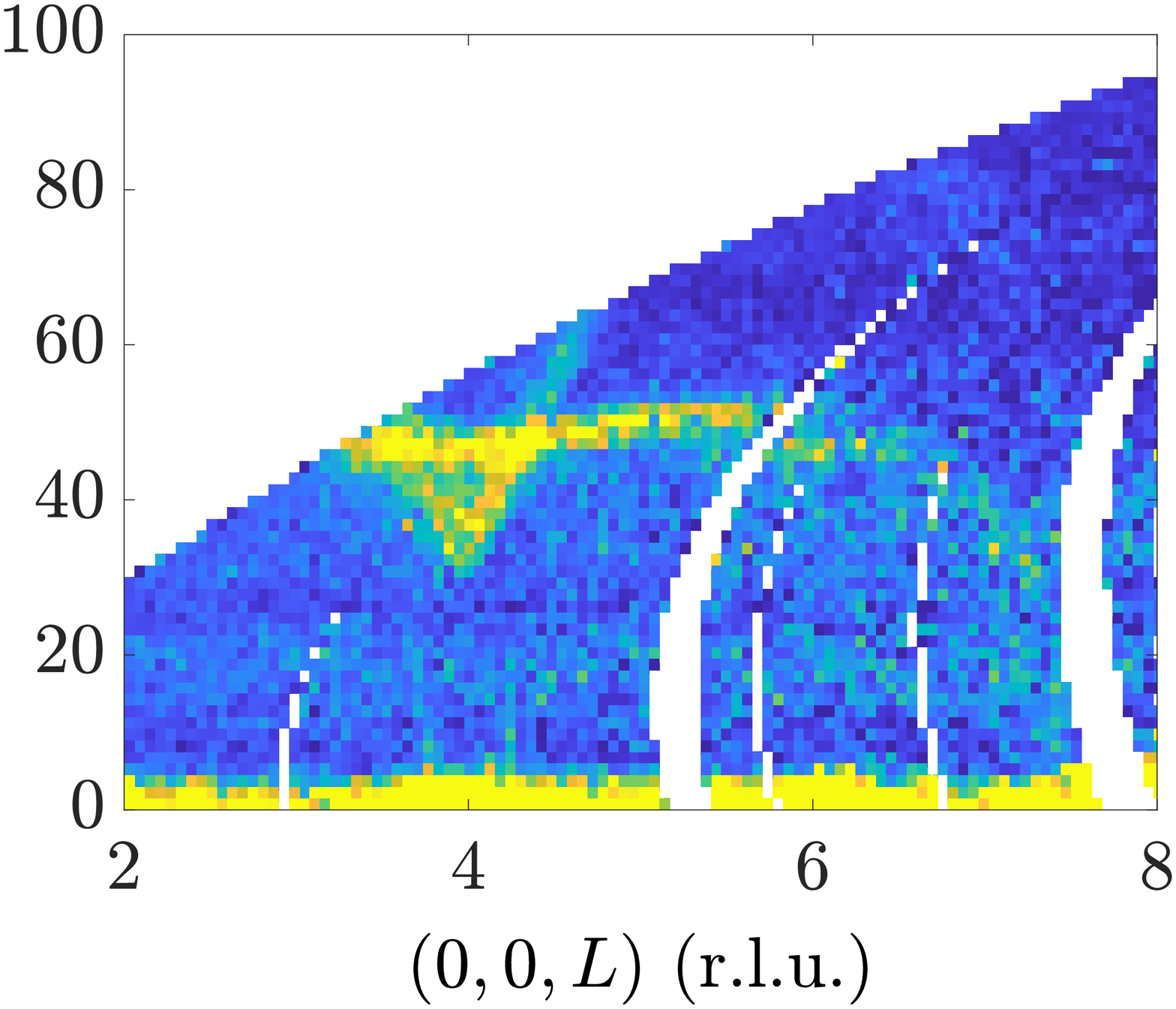}\label{fig3-b}}\vspace{-0.5em}
\subfigure{
\includegraphics[trim=0 0 0 0, clip,width=4.2cm]{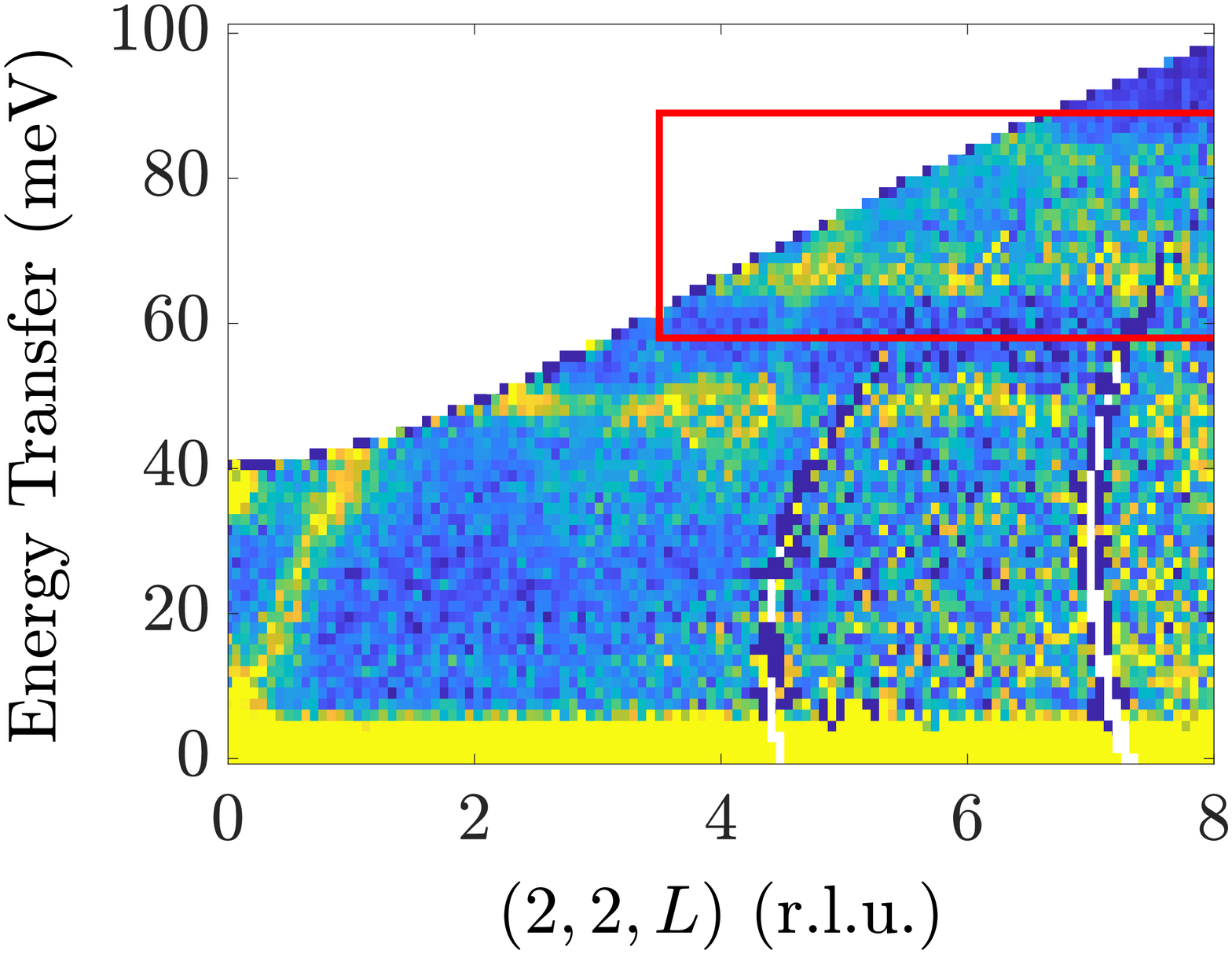}\label{fig3-c}}\hspace{-0.5em}
\subfigure{
\includegraphics[trim=0 0 0 0, clip,width=4.2cm]{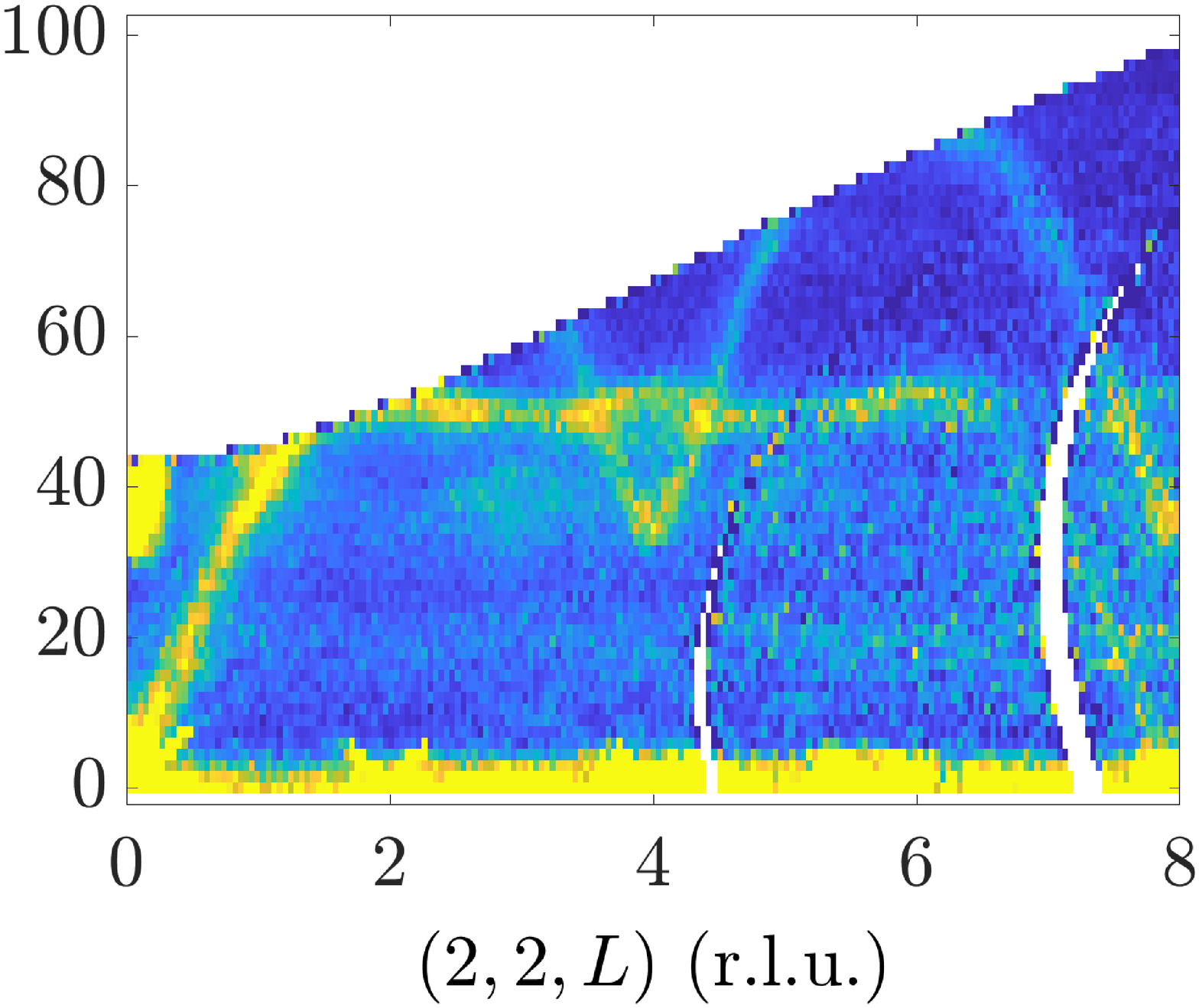}\label{fig3-d}}\vspace{-0.5em}
\caption{Data measured on MAPS at $\sim6$\,K for Yb$_3$Fe$_5$O$_{12}$ (left) and Y$_3$Fe$_5$O$_{12}$ (right). Note the similarity between the dispersive Fe spin-wave modes in both compounds. Yb single-ion excitations in YbIG are highlighted by red rectangles in the lower left panels.}\label{fig3}
\end{figure}

Figure~\ref{fig3} presents examples of neutron scattering intensity maps recorded on the MAPS spectrometer under the same conditions for Yb$_3$Fe$_5$O$_{12}$ (left panels) and Y$_3$Fe$_5$O$_{12}$ (right, data from Ref.~\onlinecite{Princep}). Differences in the signal-to-noise ratio in both data sets are caused by the different sample mass used on the YIG (12\,g) and YbIG (3\,g) experiments. The strongly dispersive Fe spin-wave modes evident in these spectra are seen to be very similar for both compounds. Already thoroughly explored in earlier works on YIG \cite{Princep,PhysRevB.97.054429}, the Fe spin-wave frequencies can be described by a set of Fe--Fe exchange interactions, of which the antiferromagnetic interaction between the two Fe$^{3+}$ sublattices is the largest (see Table~\ref{tab2-supp}). The similarity between YbIG and YIG serves as evidence that the Fe-Fe exchange is little affected by the presence of the Yb instead of Y in the garnet structure, a fortunate characteristic which is going to assist us in the determination of the energy and intensity of the CF levels in Yb$_3$Fe$_5$O$_{12}$. 

\begin{figure}
\centering
\hspace{-1em}
\subfigure{%
\includegraphics[trim=0 0 0 0, clip,width=4.25cm]{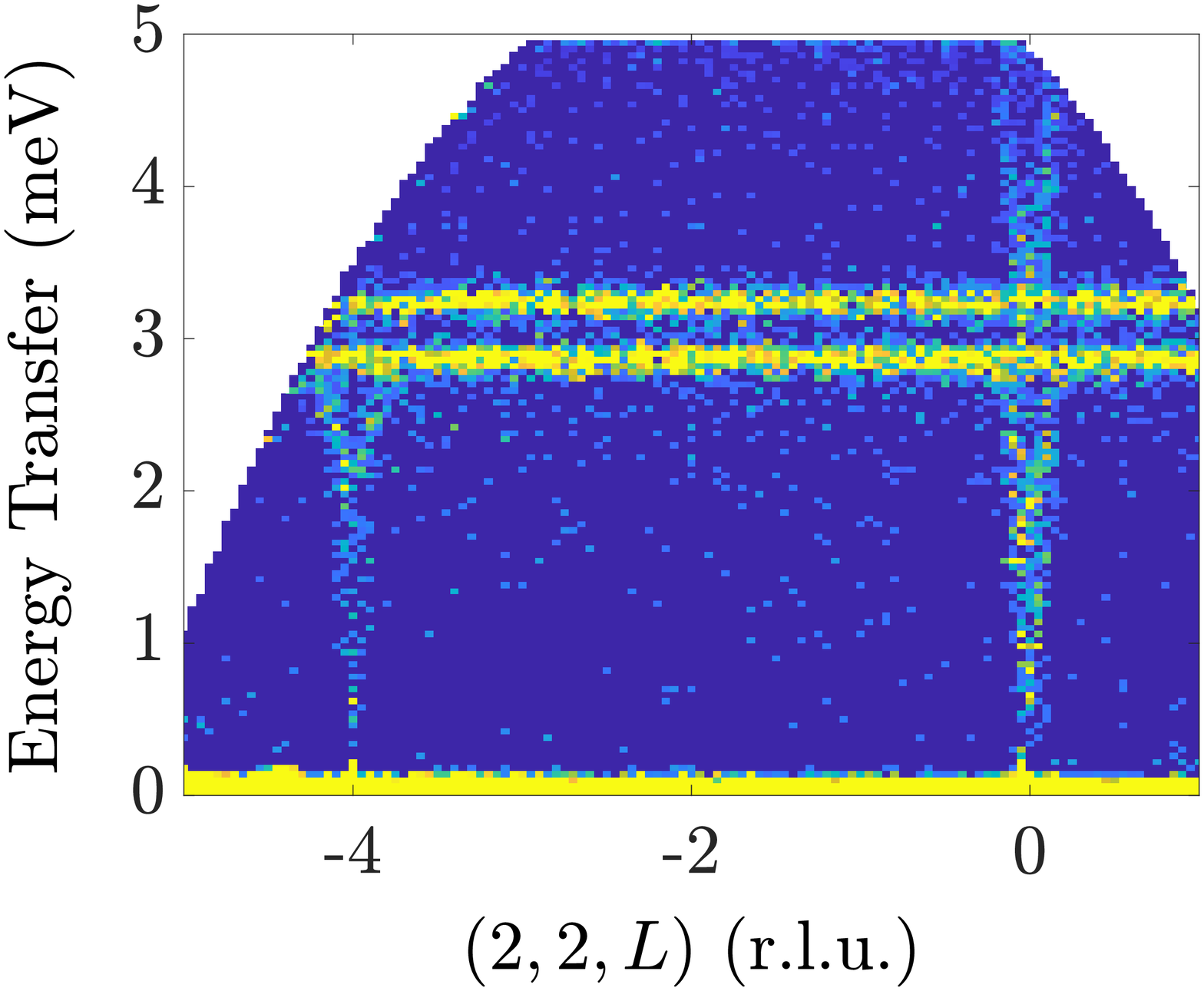}\label{fig4-a}}\hspace{0.1em}
\subfigure{%
\includegraphics[trim=40 0 0 0, clip,width=4.04cm]{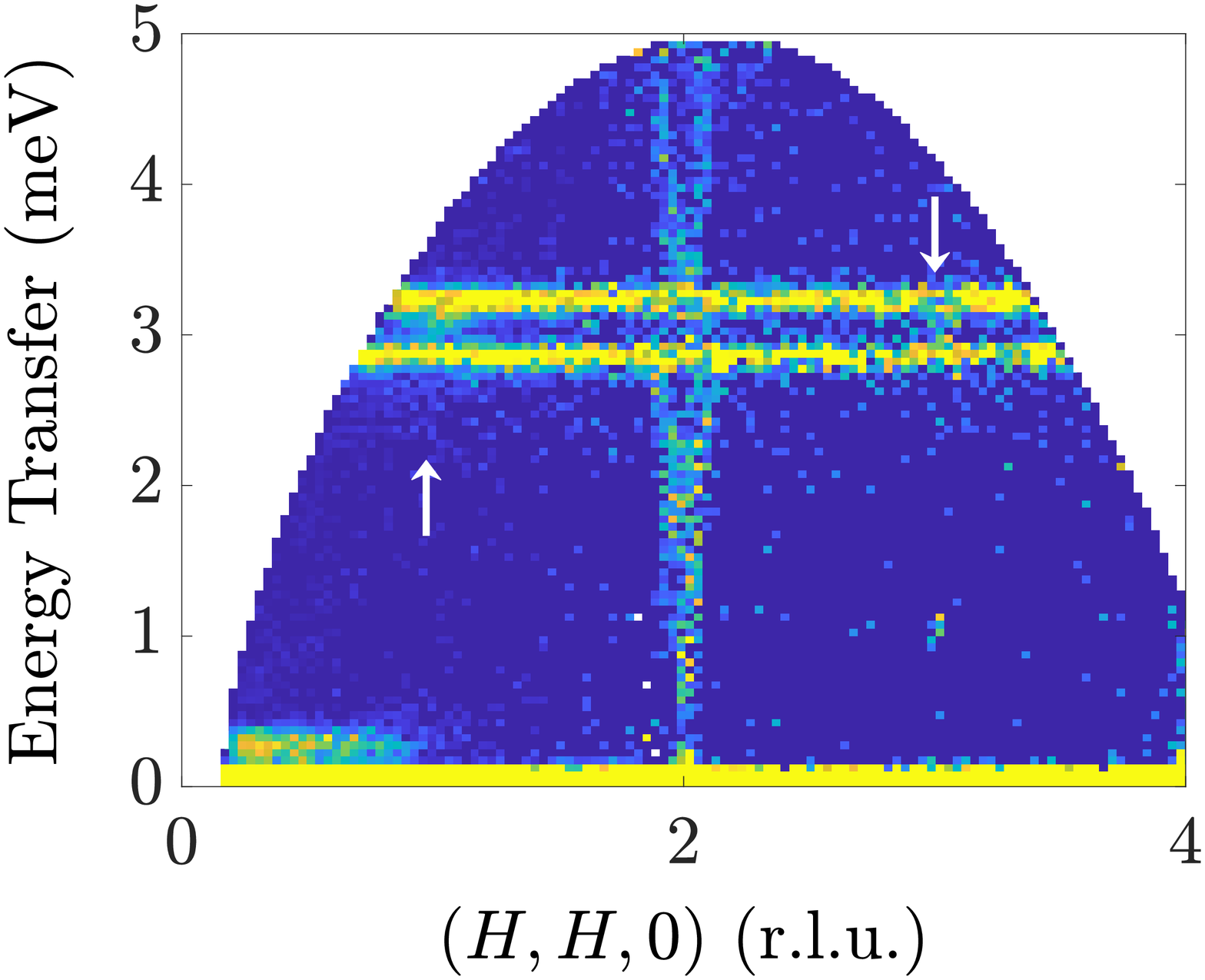}\label{fig4-b}}\vspace{-1em}
\caption{Low-energy inelastic neutron scattering data measured on LET at 1.8\,K showing the ground-state splitting of the CF excitations of the Yb$^{3+}$ ions. White arrows on the right panel highlight regions where zero intensity dispersions of Fe magnons cross the CF levels.}\label{fig4}
\end{figure}

Also in Fig.~\ref{fig3}, the bottom four panels display an extended energy interval, in which single-ion excitations of the Yb ions may be identified between 60 and 90 meV (highlighted by red rectangles), amidst some lower intensity Fe spin-wave branches.
From a neutron scattering perspective, the separation of individual single-ion levels appearing above 60 meV (see Fig.~\ref{fig2}) is challenging for two main reasons. First, these levels coincide in energy with some optical modes of the Fe spin-wave spectrum. Second, the exchange splitting of the levels is comparable with the best resolution achievable on the MAPS experiment, $\Delta E\approx 2.8$\,meV (FWHM). On the other hand, low energy inelastic neutron scattering may be used to measure the single-ion ground-state doublet splitting with much higher resolution. In Fig.~\ref{fig4}, a sampling of our LET data for energy transfers below 5\,meV is shown. Pure single-ion excitations manifest themselves as dipersionless, horizontal lines of scattering along all directions in reciprocal space. Two such levels are clearly visible in Fig.~\ref{fig4}, at energies close to 3\,meV. These correspond to the slightly different doublet splittings associated with the two symmetrically distinct Yb sites. 

Figure~\ref{fig4} also shows a typical manifestation of the hybridisation between the Yb levels and the Fe magnons. Close inspection at magnetic Brillouin zone centres, such as $(2,2,0)$ and $(2,2,-4)$, reveals the existence of a small gap between the elastic line and the softening acoustic modes (see also Fig.~\ref{fig7}). At energies closer to the CF levels, the steeply dispersing Fe spin waves split in wavevector and their slope decreases just below the Yb excitations. 

Other notable features may be seen at positions where the Fe magnons in YIG have zero structure factor, such as $(1,1,0)$ and $(3,3,0)$ \cite{Princep}. Neutron diffraction experiments on the RE iron garnets have detected, below $T_\mathrm{c}$, magnetic Bragg scattering at those reciprocal lattice wavevectors, caused by the canting of the rare-earth magnetic moments around the Fe magnetisation $\langle 111 \rangle$ directions \cite{Ghanathe,Guillot,Hock1990,Guillot1984}. In the YbIG spectrum, some intensity modulation around $(1,1,0)$ and $(3,3,0)$ may be observed at energies between the Yb levels. This may be understood as another manifestation of the Yb--Fe coupling.

\section{The model}

In order to quantitatively determine how the rare-earth single-ion transitions excite Fe magnons (and \emph{vice versa}), we develop in this section a bosonic model which can be used to treat and explain this coupling at sufficiently low temperatures. In close analogy with the conventional Holstein-Primakoff transformation, the angular momenta of the rare-earth spins are defined in terms of pseudoboson raising and lowering operators. The main difference between linear spin-wave theory and the model developed below is that, instead of reducing the coupled RE-Fe spin dynamics to that of effective spin-$\tfrac{1}{2}$ moments, several single-ion transitions of the Yb are allowed to interact with the Fe spins. Additionally, this model has the advantage of incorporating the crystal-field anisotropy directly into the exchange Hamiltonian, without the necessity of using an effective g-tensor.

\subsection{The complete Hamiltonian}

The minimal model Hamiltonian for YbIG contains the single-ion terms of Yb$^{3+}$, the exchange interaction between Fe ions, and the exchange interactions between Yb and its Fe nearest neighbours:
\begin{equation}
\mathcal{H}=\sum_{k} (\mathcal{H}^\textup{CF}_{k} + \mathcal{H}^\textup{SO}_{k}) + \sum_{\langle ij\rangle} \mathcal{H}^\textup{Ex}_{ij} + \sum_{\langle jk\rangle}\mathcal{H}^\textup{Ex}_{jk},
\label{eq6-supp}
\end{equation}
where $i,j$ are indices representing Fe sites and $k$ represents a Yb site. Note that no Yb--Yb interaction term is included in Eq.~(\ref{eq6-supp}). This is because our experiment could not detect any dispersion in the Yb single-ion excitations, as may be seen in Fig.~\ref{fig4}. 

\subsection{Crystal field Hamiltonian and symmetry considerations}

In YbIG, each one of the twelve Yb$_k$ atoms in the primitive unit cell are intercepted by three \emph{local} $2$-fold axes, which we denote here $\boldsymbol{\xi}_{k},\boldsymbol{\eta}_{k},\boldsymbol{\zeta}_{k}$ \footnote{We follow the convention established in Ref.~\onlinecite{Hutchings}}. Those are the principal axes of the crystalline electric field Hamiltonian, and do not coincide with the \emph{global}, conventional cubic axes $\boldsymbol{a},\boldsymbol{b},\boldsymbol{c}$. 
As an example of the coordinate transformation carried out in this work, for an Yb ion labelled Yb$_1$, located at the fractional coordinates $(0,3/4,7/8)$ of the conventional unit cell, we define 
\begin{align}
\begin{pmatrix}
\boldsymbol{\xi}_1\\
\boldsymbol{\eta}_1 \\
\boldsymbol{\zeta}_1 
\end{pmatrix}
=
\begin{pmatrix}
0 \ & \ 0 \ & \ 1 \\
-\frac{1}{\sqrt{2}} \ & \ -\frac{1}{\sqrt{2}} \ & \ 0 \\
\frac{1}{\sqrt{2}} \ & \ -\frac{1}{\sqrt{2}} \ & \ 0
\end{pmatrix}
\begin{pmatrix}
\mathbf{a}\\
\mathbf{b}\\
\mathbf{c}
\end{pmatrix},
\label{1}
\end{align}
where $\mathbf{a},\mathbf{b},\mathbf{c}$ are unit vectors along $\boldsymbol{a},\boldsymbol{b},\boldsymbol{c}$. The transformations for the additional Yb$_k$ $(k=2,\cdots,12)$ atoms in the primitive unit cell may be found by applying the symmetry operations of the $Ia\bar{3}d$ space group to the matrix on the right-hand side of Eq.~(\ref{1}) (see Appendix~\ref{Yb-local}). 

In the local coordinate frame defined above, $\mathcal{H}^\textup{CF}$ is identical for all the Yb$_k$ atoms, and may be written 
\begin{eqnarray}
\mathcal{H}^\textup{CF} = && B_0^2\hat{C}_0^2+B_2^2(\hat{C}_2^2+\hat{C}_{-2}^{2}) \nonumber \\
&&+B_0^4\hat{C}_0^4+B_2^4(\hat{C}_2^4+\hat{C}_{-2}^{4})+B_4^4(\hat{C}_4^4+\hat{C}_{-4}^{4}) \nonumber \\
&&+B_0^6\hat{C}_0^6 +B_2^6(\hat{C}_2^6+\hat{C}_{-2}^6)+B_4^6(\hat{C}_4^6+\hat{C}_{-4}^6) \nonumber \\
&&+B_6^6(\hat{C}_6^6+\hat{C}_{-6}^6),
\label{eq1}
\end{eqnarray}
where $\hat{C}_{\pm m}^l$ are the Wybourne tensor operators and $B_{m}^l$ are their corresponding crystal-field parameters. 

\subsection{Exchange coupling of the Fe$^{3+}$ sublattices}

The term $\mathcal{H}^\textup{Ex}_{ij}$ depends on the Fe--Fe exchange interactions describing the magnetic order and dynamics of the Fe spins. As far as our MAPS neutron scattering experiment reveals, the Fe magnon spectrum of YIG and YbIG are virtually indistinguishable (see Fig.~\ref{fig3}). Therefore, the exchange parameters reported previously for YIG \cite{Princep,PhysRevB.97.054429} are kept unchanged in our model, which may be expressed as
\begin{equation}
\mathcal{H}^\textup{Ex}_{ij}= \mathcal{J}_{ij}\,\mathbf{S}_i \cdot \mathbf{S}_j. \label{eq1-supp}
\end{equation}
We consider isotropic exchange parameters $\mathcal{J}_{ij}$ up to sixth nearest neighbours, $\mathcal{J}_1,...,\mathcal{J}_6$, and include Fe atoms on both $16a$ and $24d$ Wyckoff positions. A summary of the parameters considered in our work is given in Table~\ref{tab2-supp}.

\begin{table}
\begin{ruledtabular}
\begin{tabular}{l c c r}
$\mathcal{J}_\gamma$ & Ref.~\onlinecite{Princep} & Ref.~\onlinecite{PhysRevB.97.054429} & Coupling sites \\
\colrule
$\mathcal{J}_{1}$ & 6.8 meV & 5.80 meV & $16a$ -- $24d$ \\
$\mathcal{J}_{2}$ & 0.52 meV & 0.70 meV & $24d$ -- $24d$\\
$\mathcal{J}_{3}$ & 1.1 meV \footnote{Value quoted as $J_{3b}$ in Ref.~\cite{Princep}.} & 0.0 & $16a$ -- $16a$ \\
$\mathcal{J}_{4}$ & -0.07 meV & 0.0 & $16a$ -- $24d$ \\
$\mathcal{J}_{5}$ & 0.47 meV &- & $24d$ -- $24d$ \\
$\mathcal{J}_{6}$ & -0.09 meV &- & $16a$ -- $16a$ \\
\end{tabular}
\end{ruledtabular}
\caption{Exchange parameters obtained in Refs.~\onlinecite{Princep} and \onlinecite{PhysRevB.97.054429}, used in our Fe--Fe exchange Hamiltonian parametrisation.}\label{tab2-supp}
\end{table}
 
A transformation of the spin components is defined, so that the local ${S}^z$ of the Fe spins lies along the ordered moment direction. The iron magnetic moments on $16a$ and $24d$ sites point along any of the 8 equivalent cubic $\langle 111 \rangle$ directions, keeping always an intra-site parallel and an inter-site antiparallel orientation within a given domain. For example, if we assume that the spins of the Fe atoms in the $16a$ positions point along $[111]$, then the spins of the Fe atoms in the $24d$ positions point along $[\bar{1}\bar{1}\bar{1}]$. As long as $x,y\perp z$, the local $x$ and $y$ axes of the Fe spins can be chosen at will. Our particular choice of axes is described in Appendix~\ref{exchangeFe-Fe}.

\subsection{Yb-Fe interaction}

\begin{figure}
\centering
\includegraphics[trim=200 80 200 80, clip,width=0.35\textwidth]{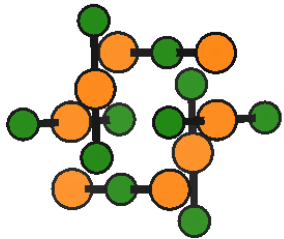}
\caption{Expanded view of the $24c$ (orange, Yb) and $24d$ (green, Fe) sublattices. Each Yb has two nearest neighbour Fe at $24d$ positions. Only bonding atoms are shown.}\label{fig1-d}
\end{figure}

The main focus of this work is on the exchange interaction $\mathcal{H}^\textup{Ex}_{jk}$. Each rare-earth in the garnet structure has two first nearest neighbour $24d$ irons, as shown in Fig.~\ref{fig1-d}. Neglecting interactions beyond these, the Yb--Fe exchange can be expressed as
\begin{equation}
\mathcal{H}^\textup{Ex}_{jk} = \mathbf{S}_j^\mathrm{T} \cdot \mathbf{A} \cdot \mathbf{J}_k,
\label{HH}
 \end{equation}
where $\mathbf{A}$ is the exchange coupling matrix. If $\mathbf{S}_j$ and the total angular momentum operator $\mathbf{J}_k$ for a particular Yb site are expressed in terms of the local coordinates of the ion Yb$_k$, then
\begin{align}
\mathbf{A}=x
\begin{pmatrix}
A^{\xi\xi} \ & \ 0 \ & \ 0 \\
0 \ & \ A^{\eta\eta}\ & 0 \\
0 \ & \ 0 \ & \ A^{\zeta\zeta} \\
\end{pmatrix}.
\label{eq5}
\end{align}
In this work, we assume that $A^{\xi\xi} \neq A^{\eta\eta} \neq A^{\zeta\zeta}$, since early optical spectroscopic data on YbIG suggested that the coupling between Yb$^{3+}$ and Fe$^{3+}$ is anisotropic \cite{PhysRevLett.4.123,PhysRev.122.1376}.

The Hamiltonian $\mathcal{H}^\textup{Ex}_{jk}$ may be split into a term $\mathcal{H}_{k}^{\textup{Ex,1}}$, which may be identified as the potential of the Yb ion in the Fe exchange field, and a dynamic coupling term $\mathcal{H}^{\textup{Ex,2}}$,
\begin{eqnarray}
\mathcal{H}^\mathrm{Ex}_{jk} =&& \mathcal{H}_{k}^{\textup{Ex,1}} +\mathcal{H}_{jk}^{\textup{Ex,2}} \nonumber \\
=&& \langle \mathbf{S} \rangle^\mathrm{T} \cdot \mathbf{A} \cdot \mathbf{J}_k + \Delta\mathbf{S}^\mathrm{T}_j \cdot \mathbf{A} \cdot \mathbf{J}_k ,
\label{eq6}
\end{eqnarray}
where $\Delta\mathbf{S}_j=\mathbf{S}_j-\langle \mathbf{S} \rangle$. We assume that $\langle \mathbf{S} \rangle$ has the fully saturated value of $S$ along the iron magnetisation. If expressed in the local coordinates of the Fe spins, $\langle \mathbf{S} \rangle=(0,0,S)$, where $S=5/2$.

\subsection{The exchange field interaction $\mathcal{H}_k^\textup{Ex,1}$}

Once the first term on the second row of Eq.~(\ref{eq6}) is expanded, as exemplified in the Appendix~\ref{exchangeYb-Fe}, it can be shown that, for the six atoms belonging to what we are going to refer as `group 1', the exchange field acting on the rare-earth is
\begin{equation}
\mathcal{H}_{k}^{\textup{Ex,1}}=-\frac{nS}{3}(\sqrt{3}A^{\xi\xi}J_k^{\xi}-\sqrt{6}A^{\eta\eta}J_k^{\eta}),
\label{eq2}
\end{equation}
while, for atoms of the `group 2'
\begin{equation}
\mathcal{H}_{k}^{\textup{Ex,1}}= \frac{nS}{3}(\sqrt{3}A^{\xi\xi} J_k^{\xi}-\sqrt{6}A^{\zeta\zeta} J _k^{\zeta}),
\label{eq3}
\end{equation}
where $n=2$ is the number of nearest neighbours of the Yb$^{3+}$. 

A part depending only on the rare-earth index $k$ can now be separated from the Hamiltonian in Eq.~(\ref{eq6-supp}) 
\begin{equation}
\mathcal{H}_1=\sum_k \mathcal{H}^\textup{SO}_{k} + \mathcal{H}^\textup{CF}_{k}+\mathcal{H}_{k}^{\textup{Ex,1}},
\label{H_1}
\end{equation} 
which is, initially, diagonalised separately from $\mathcal{H}_{jk}^{\textup{Ex,2}}$. The eigenvalues $E_{km}$, with eigenfunctions $\Gamma_{km}$, of $\mathcal{H}_1$ are going to be the energies of the single-ion modes measured on MAPS and LET. 

\subsection{Quadratic form of $\mathcal{H}$}

In this second step of data analysis, we follow the formalism introduced by Grover \cite{PhysRev.140.A1944} and extensively expanded in Refs.~\onlinecite{PhysRev.167.510,Buyers_1971}. At sufficiently low temperatures, the angular momentum operator $\mathbf{J}_k$, up to first order in the pseudoboson operators $c^\dagger_{km},c_{km}$ (see defintion in Appendix~\ref{exchange_hamiltonian}), for a level $m$ may be replaced by \cite{Buyers_1971}
\begin{equation}
\mathbf{J}=\mathbf{J}_{00} + \sum_m \mathbf{J}_{0m} c^\dagger_{m} + \mathbf{J}_{m0} c_{m}
\label{eq22}
\end{equation}
where $\mathbf{J}_{mn}=\langle \Gamma_{n} | \mathbf{L}+\mathbf{S} | \Gamma_{m} \rangle$ and the index $k$ is omitted for clarity. Now, the components of $\mathbf{J}$ in Eq.~(\ref{eq6}) can be substituted by Eq.~(\ref{eq22}). The complete Hamiltonian $\mathcal{H}$ in Eq.~(\ref{eq6-supp}) may be finally written solely in terms of pseudoboson operators
\begin{align}
\mathcal{H}=& S\Big\{ \sum_{\langle ii'\rangle} \mathcal{J}_{ii'} a_i^\dagger a_{i'} + \sum_{\langle jj'\rangle}\mathcal{J}_{jj'}d_j^\dagger d_{j'} \nonumber \\
&+ \sum_{\langle ij\rangle}(\mathcal{J}_{ij}a_{i}d_{j} + \textrm{h.c.}) \Big\}+\sum_{k,m>0} E_{km} c^\dagger_{km} c_{km} \nonumber \\
&+\sum_{\langle{jk}\rangle} (K^{(1)}_{jk} d_j^\dagger c^\dagger_{km} + K^{(2)}_{jk} d_j^\dagger c_{km} - K_{jk}^{(3)} d_j^\dagger d_j + \textup{h.c.}), 
\label{H_final}
\end{align}
where h.c. denotes hermitian conjugate, and only terms up to second order are retained. The first three summations are over pairs of Fe atoms, with the $(i, i')$ referring to atoms on the $16a$ sites, and $(j, j')$ to the $24d$ sites. The standard Holstein--Primakoff operators $(a, a^{\dag})$ and $(d,d^{\dag})$, describing Fe spin deviations on the $16a$ and $24d$ sites, respectively, are defined in Appendix~\ref{exchangeFe-Fe}. Values for the coefficients $K_{jk}^{(1)},K_{jk}^{(2)},K_{jk}^{(3)}$ are given in Appendix~\ref{exchange_hamiltonian}. 

The quadratic Hamiltonian for $N=12$ Yb ions has dimensions $2N\times2N$ for each CF level $m=1,...,7$ included in the model. The iron part contributes additionally with a $40\times40$ block to the Hamiltonian. In total, the full $\mathcal{H}$ would be a matrix with dimensions of $208\times208$. As the ground-state Kramers doublet of Yb is separated by $\sim60$\,meV from the higher excited CF states, it is a good approximation to include only the first excited CF level ($m=1$) in the calculation. The size of the matrix is then reduced, and so the time necessary for the numerical diagonalisation of $\mathcal{H}$. More details about the reduced $64\times64$ matrix representing Eq.~(\ref{H_final}) are given in Appendix~\ref{exchange_hamiltonian}. 

The calculation of the neutron scattering cross-section for the excitations observed in this work is detailed in the Appendix~\ref{Xsec}.

\section{Data Analysis}

\begin{figure}
\centering
\includegraphics[trim=0 0 0 0, clip,width=8.2cm]{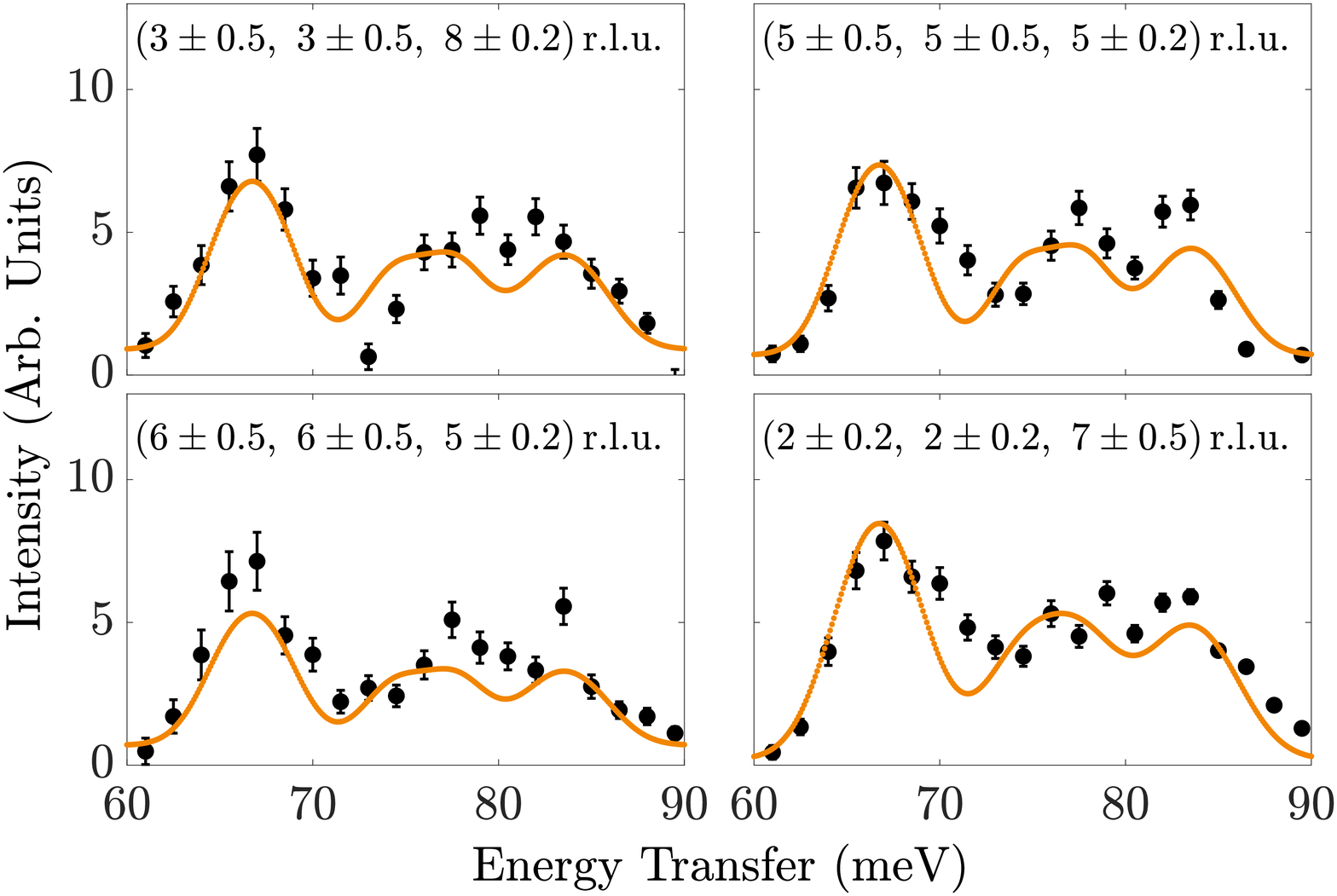}\vspace{-0.8em}
\caption{Examples of experimental scattering from CF excitations (circles, with error bars), obtained following the procedure detailed in the text. The orange line corresponds to the best fit found in this work. The spectra are averaged over the range in $\mathbf{Q}$ indicated in each panel.}\label{fig5}
\end{figure}

In order to isolate the scattering intensity of the single-ion modes in the MAPS data, identical constant-$\mathbf{Q}$ cuts were performed on the measured spectra of Yb$_3$Fe$_5$O$_{12}$ and Y$_3$Fe$_5$O$_{12}$. The latter is considered to provide an approximate background of excitations common to both compounds. Despite the difference in mass between Y and Yb, we assume that the phononic background at the relevant, medium energy range in YbIG is only negligibly different from that in YIG. Further discussion, including a raw data comparison, is given in Appendix \ref{det_MAPS}. Some examples of profiles obtained after subtraction of YIG from YbIG data are shown in Fig.~\ref{fig5}.

To obtain experimental intensities and energies, the background-subtracted data were fitted with three Gaussians, whose widths are limited by the experimental FWHM. The intensity ratios, normalised so that the sum of the integral of the three peaks are equal to unity, and peak centres are listed in Table~\ref{tab1}. A similar procedure was carried out in order to extract the energies of the dispersionless modes from the LET data. The intensities of the low-energy peaks are not are given in Table~\ref{tab1} because they could not be reliably scaled to the MAPS intensities. The spectral weight of both low energy modes is nevertheless taken into consideration in the second part of our analysis, as we will show later. Also note that, even though only the $J=7/2$ levels have energies lower than $100\ \textup{meV}$, data on the $J=5/2$ levels were also included in our parametrisation of the combined crystal field and spin-orbit Hamiltonian. The energy of the lowest level of the $J=5/2$ manifold, not accessible in our experiment, is taken from Ref. \cite{PhysRev.122.1376}. 

\begin{table}
\begin{ruledtabular}
\begin{tabular}{l c c c c}
 	 & \multicolumn{2}{c}{Energy (meV)} & \multicolumn{2}{c}{Normalised Intensity} \\
\colrule
Level & Measured & Calculated & Measured & Calculated \\
\colrule
$E^1_0$ & \multirow{2}{*}{-} & 0 & \multirow{2}{*}{-} & 0.11 \\
$E^2_0$ & & 0 & & 0.16 \\
\colrule
$E^1_1$ & $2.868\pm0.009$ & 2.881 & \multirow{2}{*}{-} & 0.25 \\
$E^2_1$ & $3.247\pm0.009$ & 3.235 & & 0.19 \\
\colrule
$E^1_2$ & \multirow{4}{*}{$67.0\pm0.3$} & 65.4 & \multirow{4}{*}{0.46} & 0.10 \\
$E^2_2$ & & 65.4 & & 0.09 \\
$E^1_3$ & & 67.4 & & 0.13 \\
$E^2_3$ & & 68.1 & & 0.11\\
\colrule
$E^1_4$ & \multirow{4}{*}{$76.4\pm1.0$} & 75.5 & \multirow{4}{*}{0.26} & 0.01 \\
$E^2_4$ & & 74.4 & & 0.15 \\
$E^1_5$ & & 77.7 & & 0.15 \\
$E^2_5$ & & 78.6 & & 0.02 \\
\colrule
$E^1_6$ & \multirow{4}{*}{$82.7\pm0.7$} & 82.1 & \multirow{4}{*}{0.28} & 0.06 \\
$E^2_6$ & & 83.1 & & 0.07 \\
$E^1_7$ & & 85.3 & & 0.06 \\
$E^2_7$ & & 84.6 & & 0.05 \\
\colrule
$E^1_8$ & \multirow{4}{*}{$1276\pm7$\footnote{Value from Ref. \cite{PhysRev.122.1376}.} } & 1275 & \multirow{4}{*}{-} & \multirow{4}{*}{-} \\
$E^2_8$ & & 1275 & & \\
$E^1_{9}$ & & 1276 & & \\
$E^2_{9}$ & & 1277 & & \\
\end{tabular}
\end{ruledtabular}
\caption{Experimental and calculated energies and intensities of the single-ion Hamiltonian. In the first column, the subscript $k$ in $E_{km}$ was replaced by a generalised superscript, either $1$ or $2$, referring to atoms of group 1 or 2, respectively. Calculated values are those corresponding to the parameters shown in the last column of Table~\ref{tab2}.}\label{tab1}
\end{table}

The iterative procedure (repeated until a best model is obtained) conducted in order to fit $\mathcal{H}$ to the data can be summarised as follows:
\begin{itemize}
\item{the Fe$^{3+}$ exchange field was initially neglected, and the spin-orbit parameter $\lambda$ and the $B_{m}^l$ in Eq.~(\ref{eq1}) were varied to give the best fit to the experimental energies and relative peak intensities;}
\item{The exchange field was added and determined from the measured ground-state splittings ($E_1^1=2.868$\,meV and $E_1^2=3.247$\,meV), corresponding to the two inequivalent Yb sites. Possible values for $(A^{\xi\xi}, A^{\eta\eta},A^{\zeta\zeta})$ are estimated using $E_1^1$ and $E_1^2$;}
\item{The full model for $\mathcal{H}$ is compared to the data, including those shown in Fig.~\ref{fig5}. The best values for $(A^{\xi\xi}, A^{\eta\eta},A^{\zeta\zeta})$ are confirmed by comparison with the experimental hybridisation features (see below).} 
\end{itemize}

In the first part of the fitting (first bullet point above), we used as starting parameters the $\lambda$ and the $B_{m}^l$ obtained for the exchange-field-free compound Yb$_3$Ga$_5$O$_{12}$ (YbGG) in Ref.~\onlinecite{PhysRev.159.251}. Given that the local oxygen environment of the Yb ions in YbGG and YbIG is very similar \cite{Hutchings}, this assumption is justifiable. In the least-squares fitting, we allowed the spin-orbit coupling and all nine crystal field parameters (see Eq.~\ref{eq1}) to vary. As observables, the experimental transition energies and intensity ratios obtained in this work (see Table~\ref{tab1}) were combined with the $^{2}F_\frac{7}{2} \rightarrow\ ^{2}F_\frac{5}{2}$ transition energies of YbIG and dilute Yb in Y$_3$Ga$_5$O$_{12}$ from the optical study reported in Ref.~\onlinecite{PhysRev.122.1376}. An additional constraint is provided by the fact that the rare-earth CF environment is approximately cubic, as demonstrated in previous studies of the rare-earth garnets \cite{PhysRev.118.1490,Hutchings,VanVleck}. This constraint allowed us to limit the parameter space and reject any fits in which the largest CF parameters, in an absolute sense, were not the cubic parameters $B_{0}^4$, $B_{4}^4$, $B_{0}^6$ and $B_{4}^6$.

\begin{figure}
\centering
\includegraphics[trim=38 0 15 15, clip,width=8.8cm]{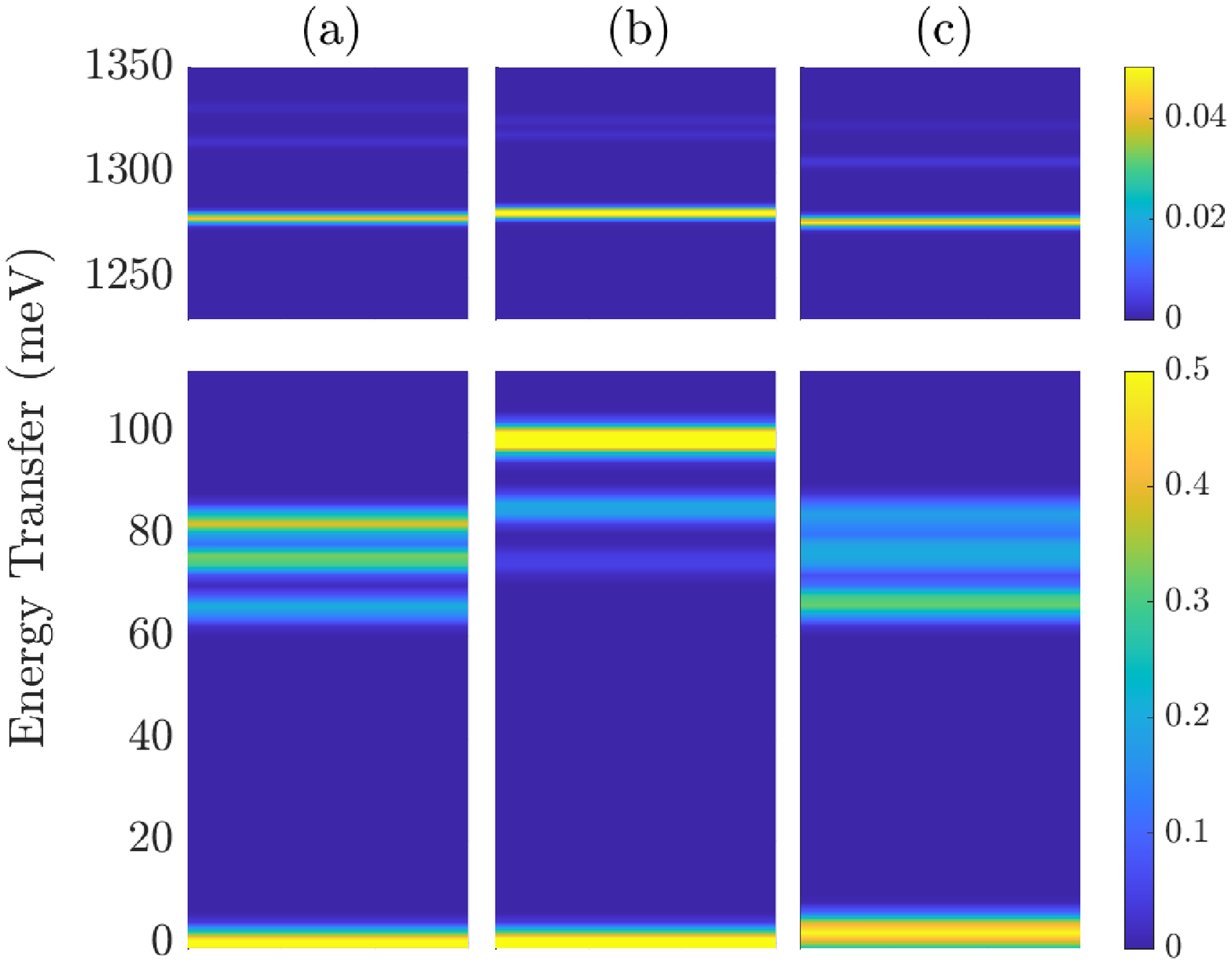}
\caption{Diagram representing energy levels and intensities at $T=5$\,K deduced for \hyperref[fig6]{(a)} $\mathrm{Yb_3Ga_5O_{12}}$ and \hyperref[fig6]{(b)} $\mathrm{Yb_3Al_5O_{12}}$ in Ref. \onlinecite{PhysRev.159.251}. In \hyperref[fig6]{(c)} the best fit for $\mathrm{Yb_3Fe_5O_{12}}$ obtained in this work is shown. The calculated intensities were convolved with a Gaussian with the instrumental resolution width. Note the different colour scale in the bottom and top panels. Intensities were calculated using the dipole approximation, and are rigorously correct only for $\mathbf{Q}=0$. }\label{fig6}
\end{figure}

The best-fit parameters are shown in the last column of Table~\ref{tab2}. For comparison, parameters for the sister compounds $\mathrm{Yb_3Ga_5O_{12}}$ and $\mathrm{Yb_3Al_5O_{12}}$, obtained in Ref. \onlinecite{PhysRev.159.251} are also shown. Note that Ga$^{3+}$ and Al$^{3+}$, unlike Fe$^{3+}$, are non-magnetic ions. As a result, the degeneracy of the Yb$^{3+}$ Kramers doublets in these compounds is not lifted. The calculated spectra for each of the models are displayed in Figs.~\hyperref[fig6]{7(a)}-\hyperref[fig6]{7(c)}, following the same order of Table~\ref{tab2}. Intensities were estimated at $T=5$\,K and convolved with Gaussians of $\textup{FWHM}=2.8\ \textup{meV}$. Constant-$\mathbf{Q}$ cuts analogous to those performed on the experimental spectra are shown along with the data in Fig.~\ref{fig5}. 

\begin{table}
\begin{ruledtabular}
\begin{tabular}{l c c r}
 Parameter& \multicolumn{2}{c}{Pearson \emph{et al.} \cite{PhysRev.159.251}} & This work \\
 (meV) & ($\mathrm{Yb_3Ga_5O_{12}}$)	& ($\mathrm{Yb_3Al_5O_{12}}$) & $(\mathrm{Yb_3Fe_5O_{12}}$) \\
\colrule
$\lambda$ & -357 & -354 & -357 \\
$B_0^2$ & -17.4 & -3.0 & -10.9\\
$B_2^2$ & 29.2 & 35.6 & 11.2\\
$B_0^4$ & -175.6 & -149.8 & -211.8 \\
$B_2^4$ & 34.7 &16.0 & 36.7 \\
$B_4^4$ & 73.6 & 82.0 & 75.3\\
$B_0^6$ & 97.2 & 212.3 &76.9 \\
$B_2^6$ & -89.4 & -148.5 &-38.5\\
$B_4^6$ & 145.3 & 178.3 &131.2\\
$B_6^6$ & -30.4 & -30.0&-39.0\\
\colrule
$A^{\xi\xi}$ & \multicolumn{3}{r}{0.137}\\
$A^{\eta\eta}$ & \multicolumn{3}{r}{0.262}\\
$A^{\zeta\zeta}$ & \multicolumn{3}{r}{0.226}
\end{tabular}
\end{ruledtabular}
\caption{Comparison between crystal-field parameters found in this work and those estimated in Ref.~\onlinecite{PhysRev.159.251} for exchange-field free compounds $\mathrm{Yb_3Ga_5O_{12}}$ and $\mathrm{Yb_3Al_5O_{12}}$.}\label{tab2}
\end{table}

\begin{figure*}
\centering
\subfigure{%
\includegraphics[trim=0 0 0 0, clip,width=6cm]{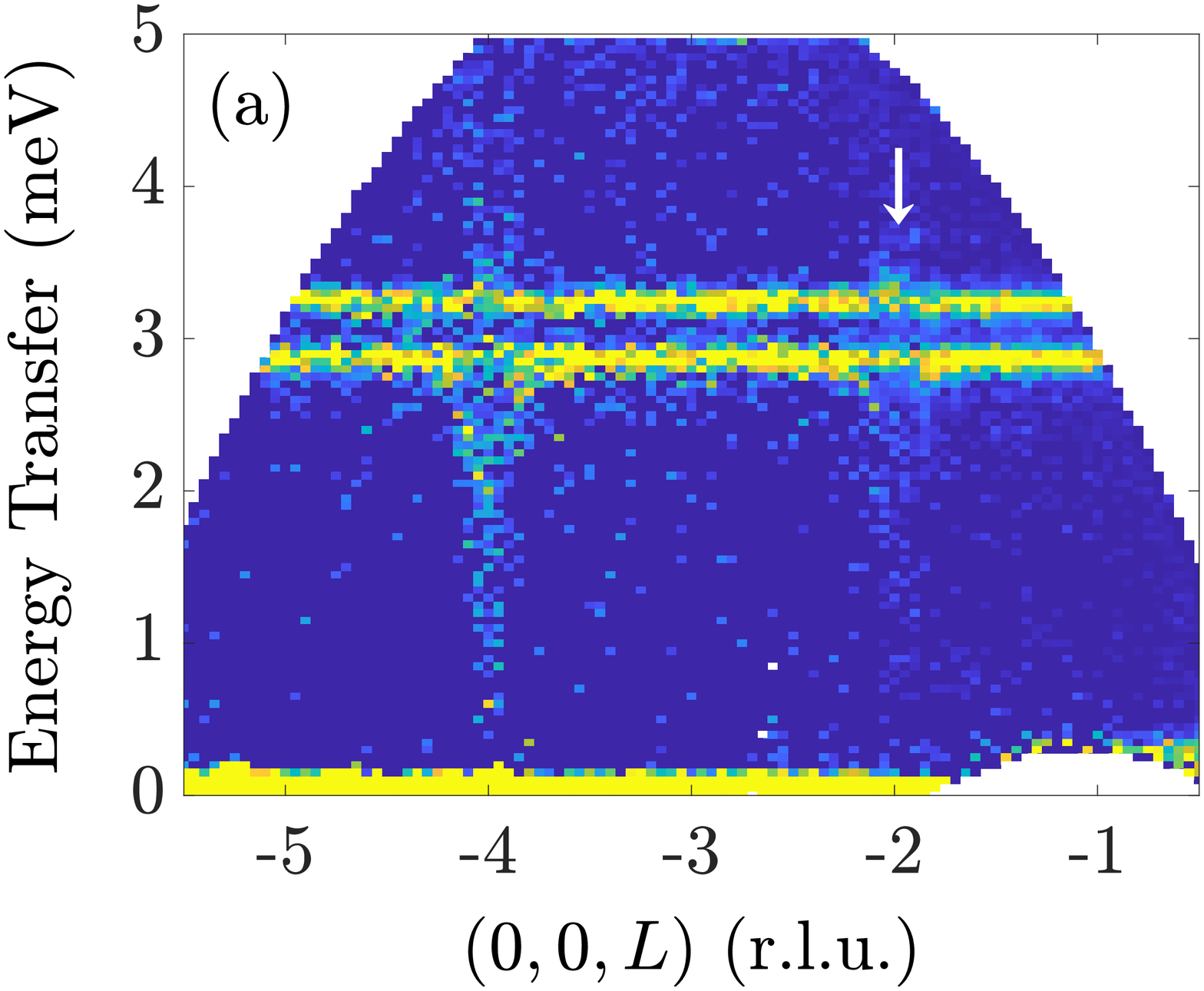}}\hspace{-0.2em} 
\subfigure{%
\includegraphics[trim=40 0 0 0, clip,width=5.7cm]{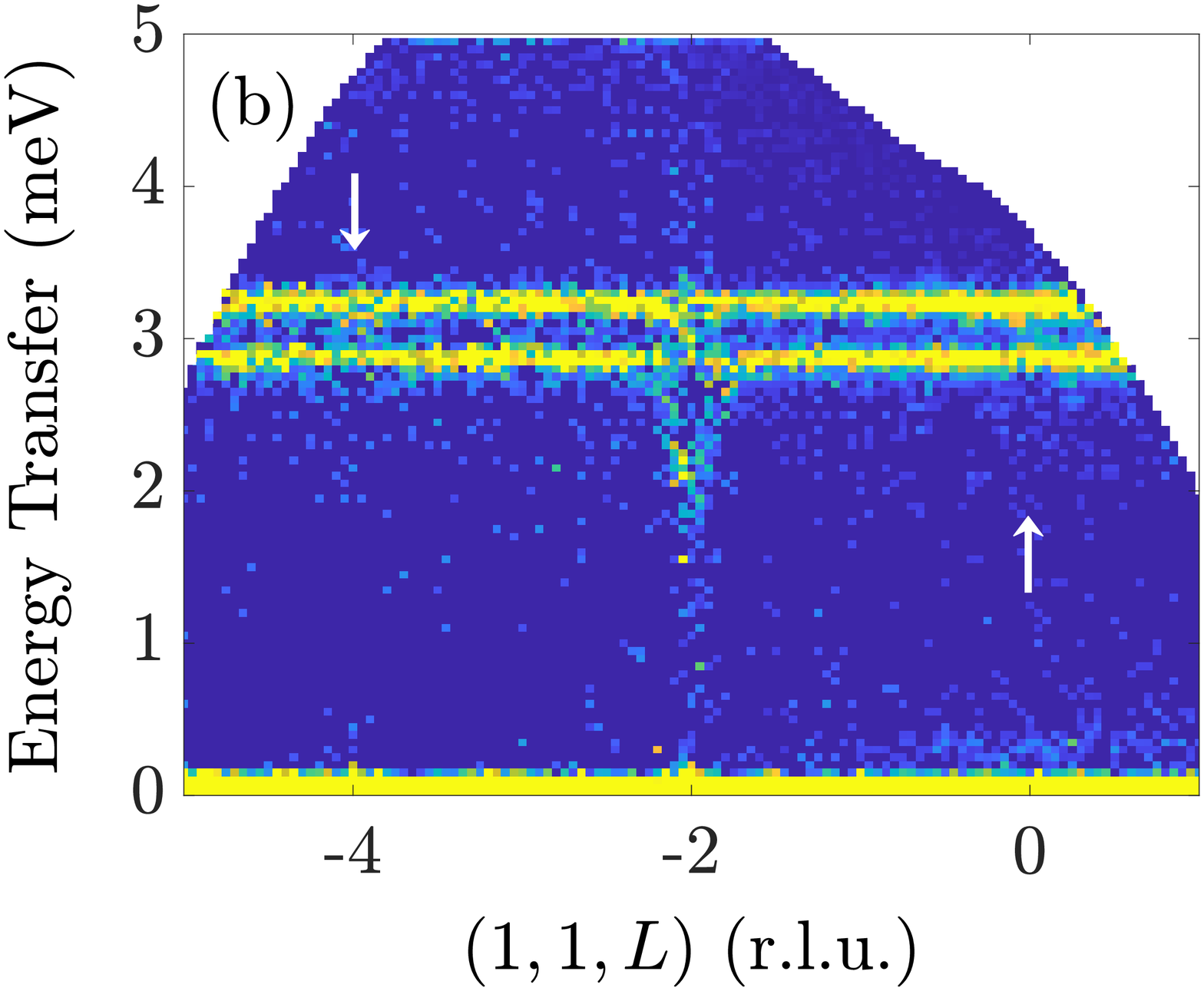}}\hspace{-0.2em} 
\subfigure{%
\includegraphics[trim=40 0 0 0, clip,width=5.7cm]{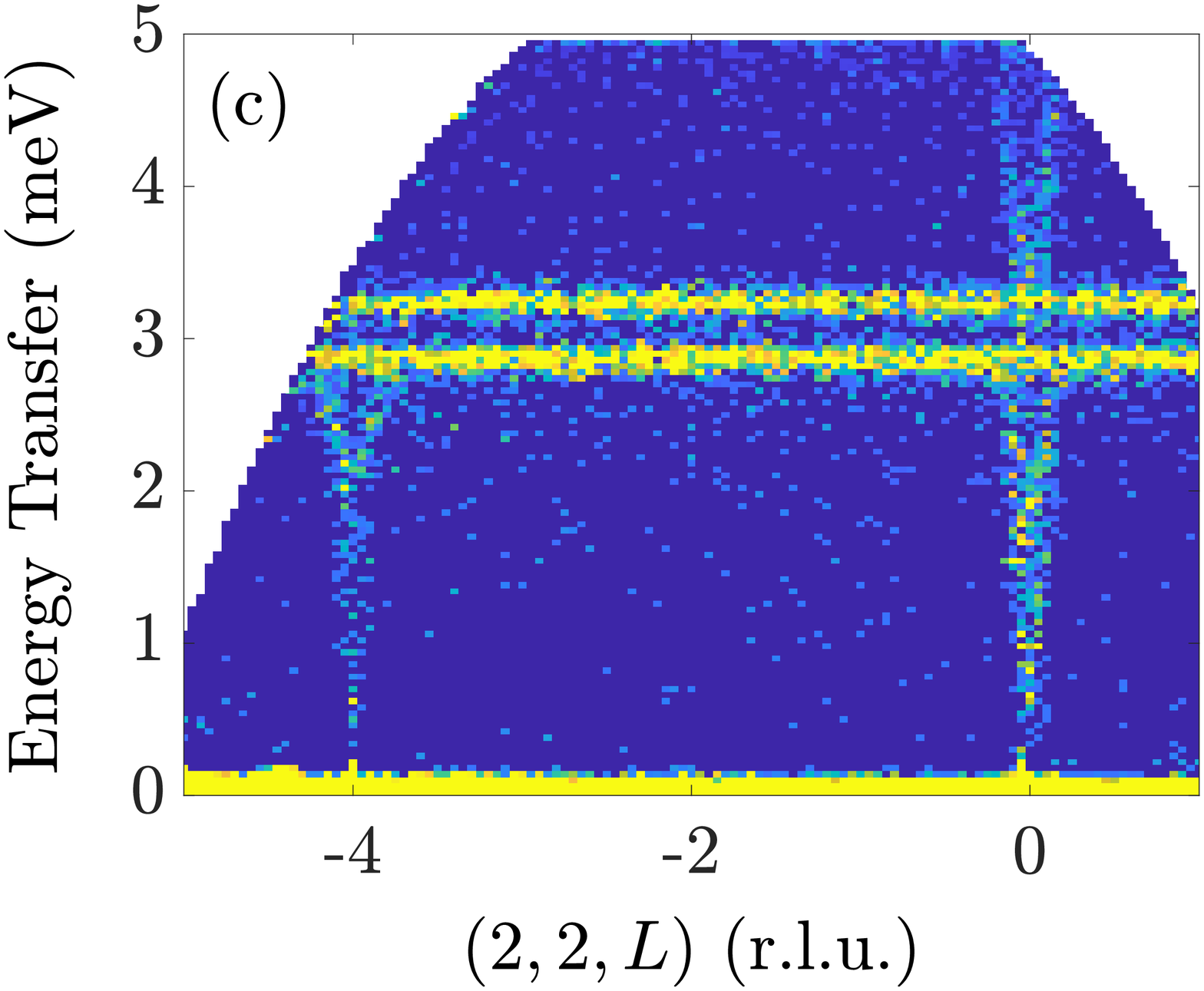}}\vspace{-2.1em}
\subfigure{%
\includegraphics[trim=0 0 30 80, clip,width=6.14cm]{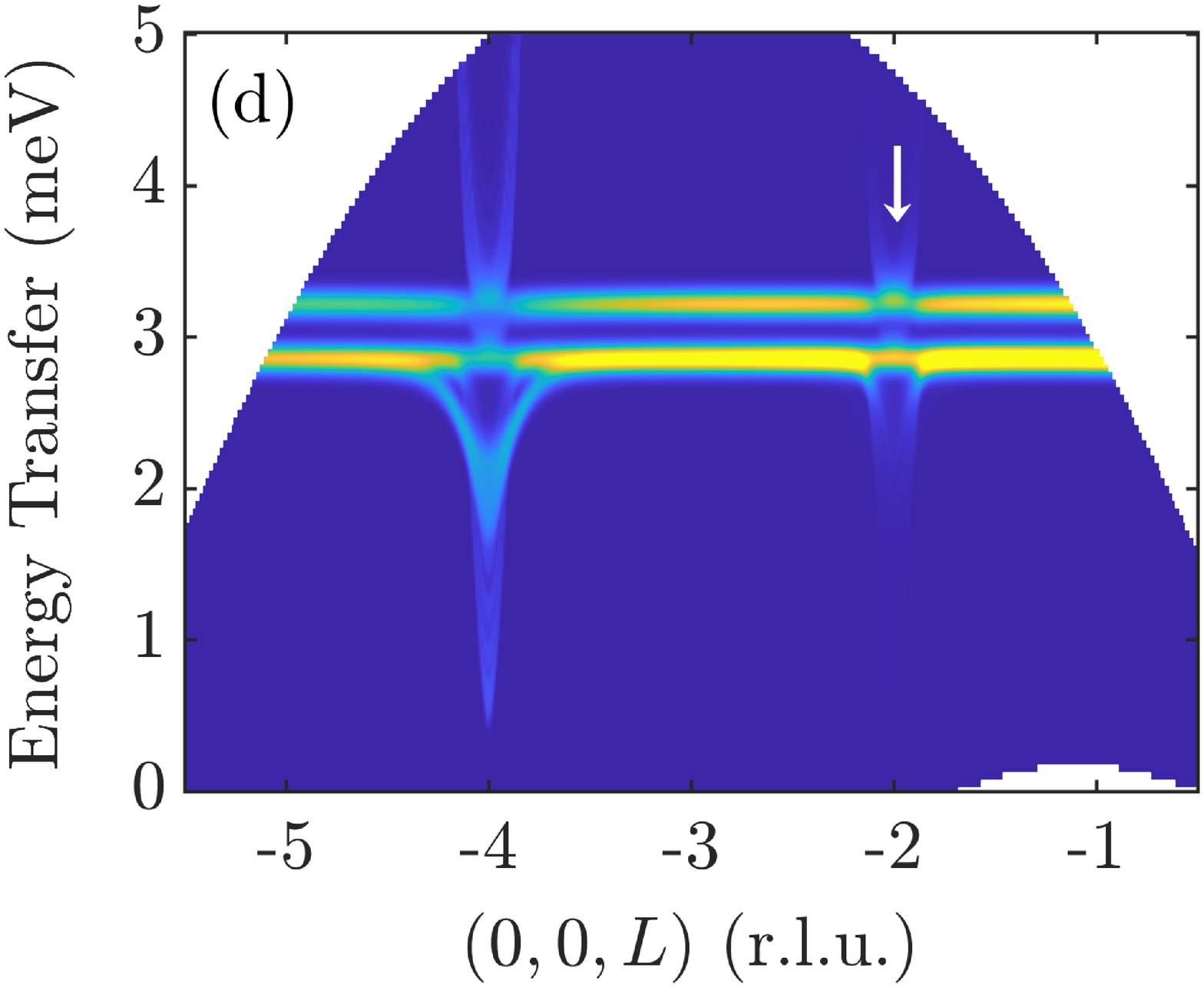}}\hspace{-0.2em} 
\subfigure{%
\includegraphics[trim=50 0 30 80, clip,width=5.7cm]{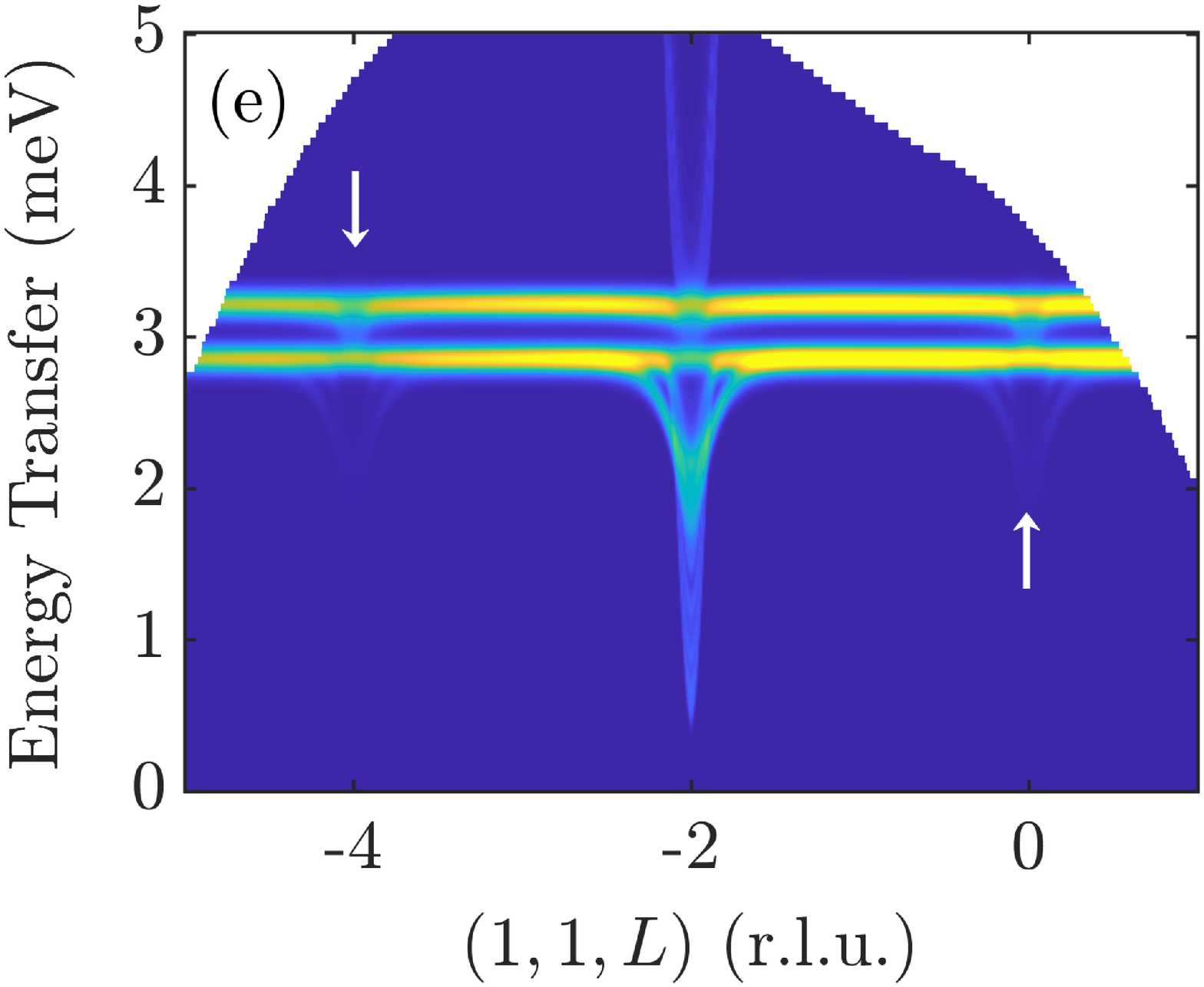}}\hspace{-0.2em} 
\subfigure{%
\includegraphics[trim=50 0 30 80, clip,width=5.7cm]{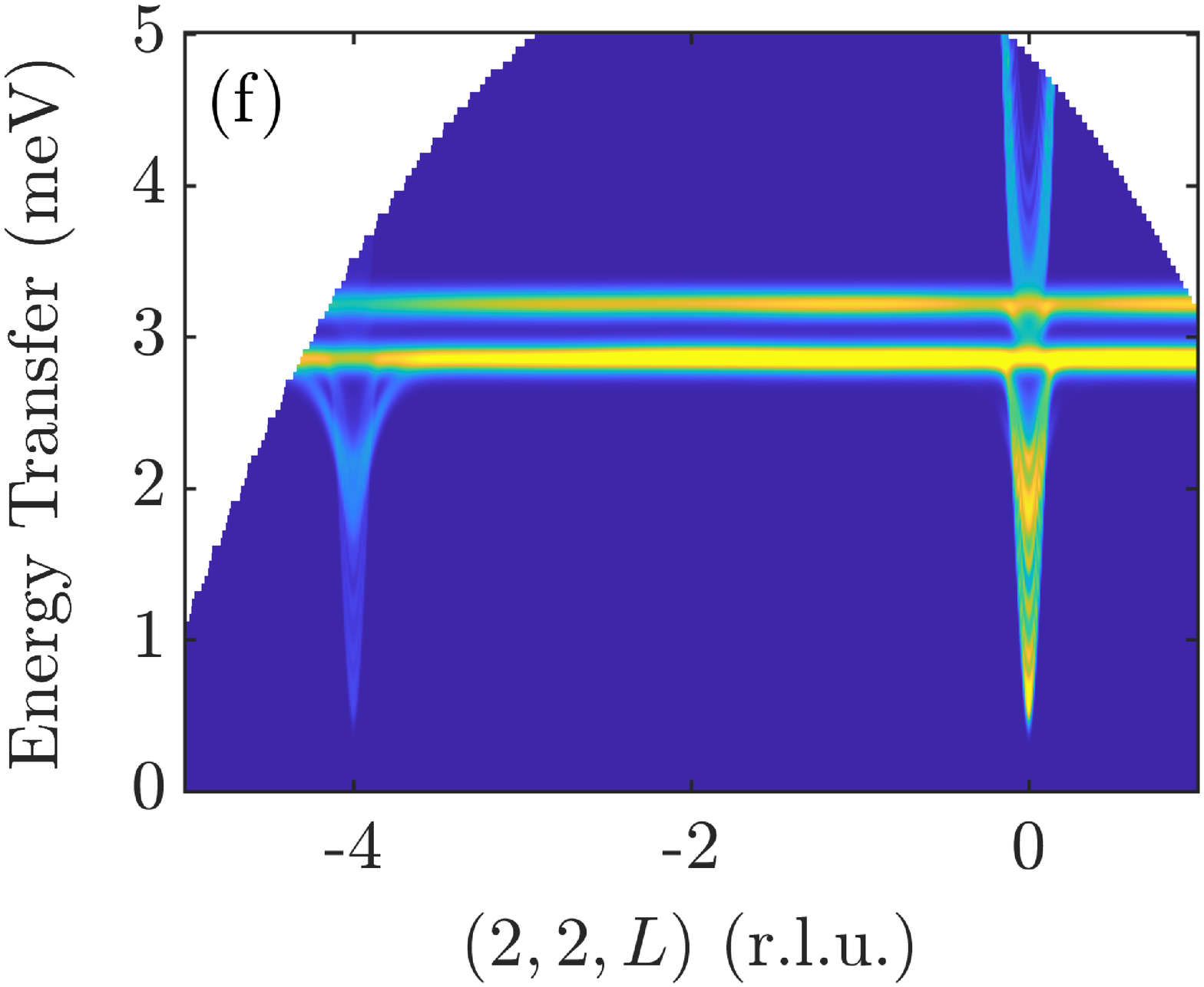}}\vspace{-2.1em} 
\subfigure{%
\includegraphics[trim=0 0 0 0, clip,width=6.12cm]{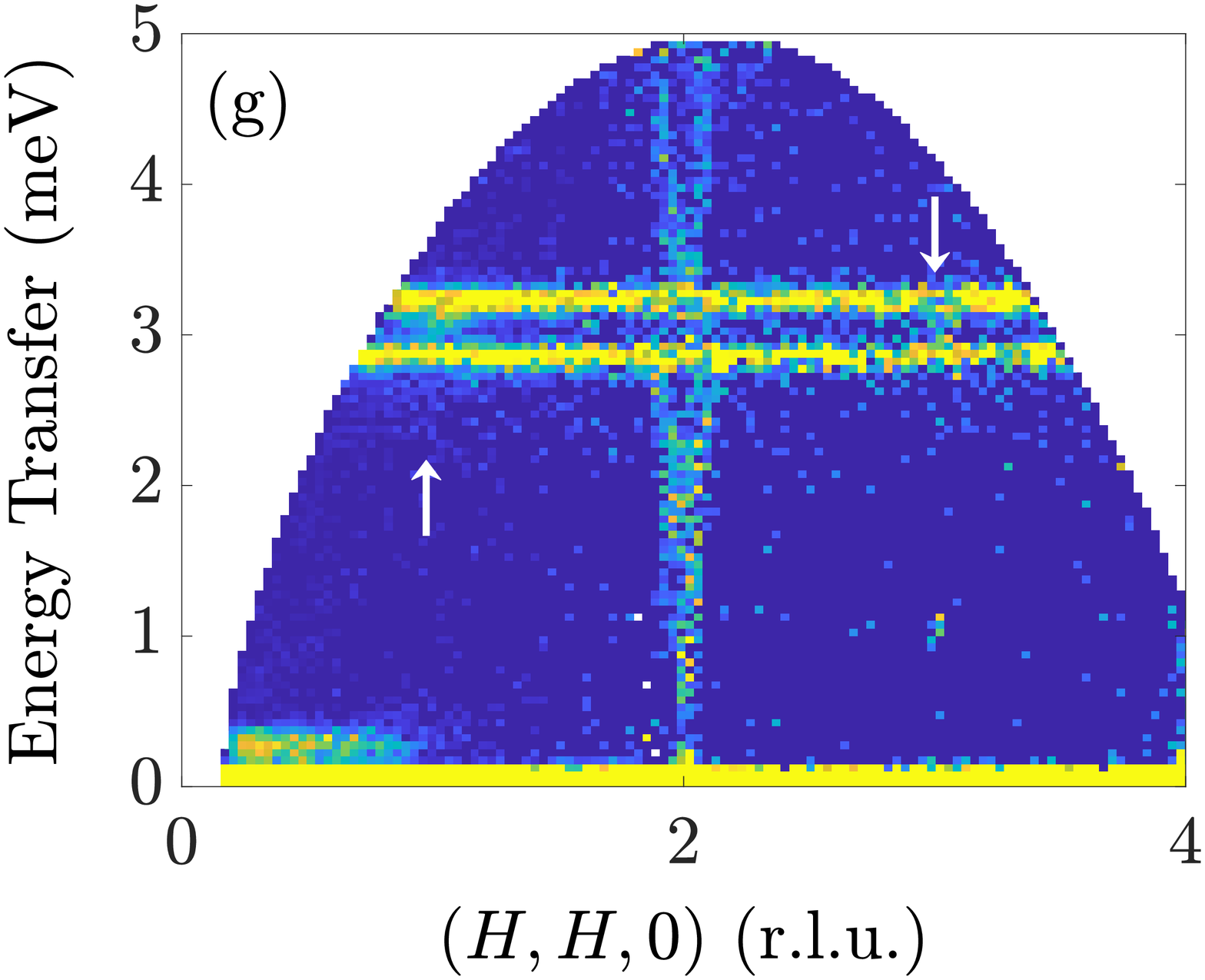}}\hspace{-0.2em}
\subfigure{%
\includegraphics[trim=40 0 0 0, clip,width=5.7cm]{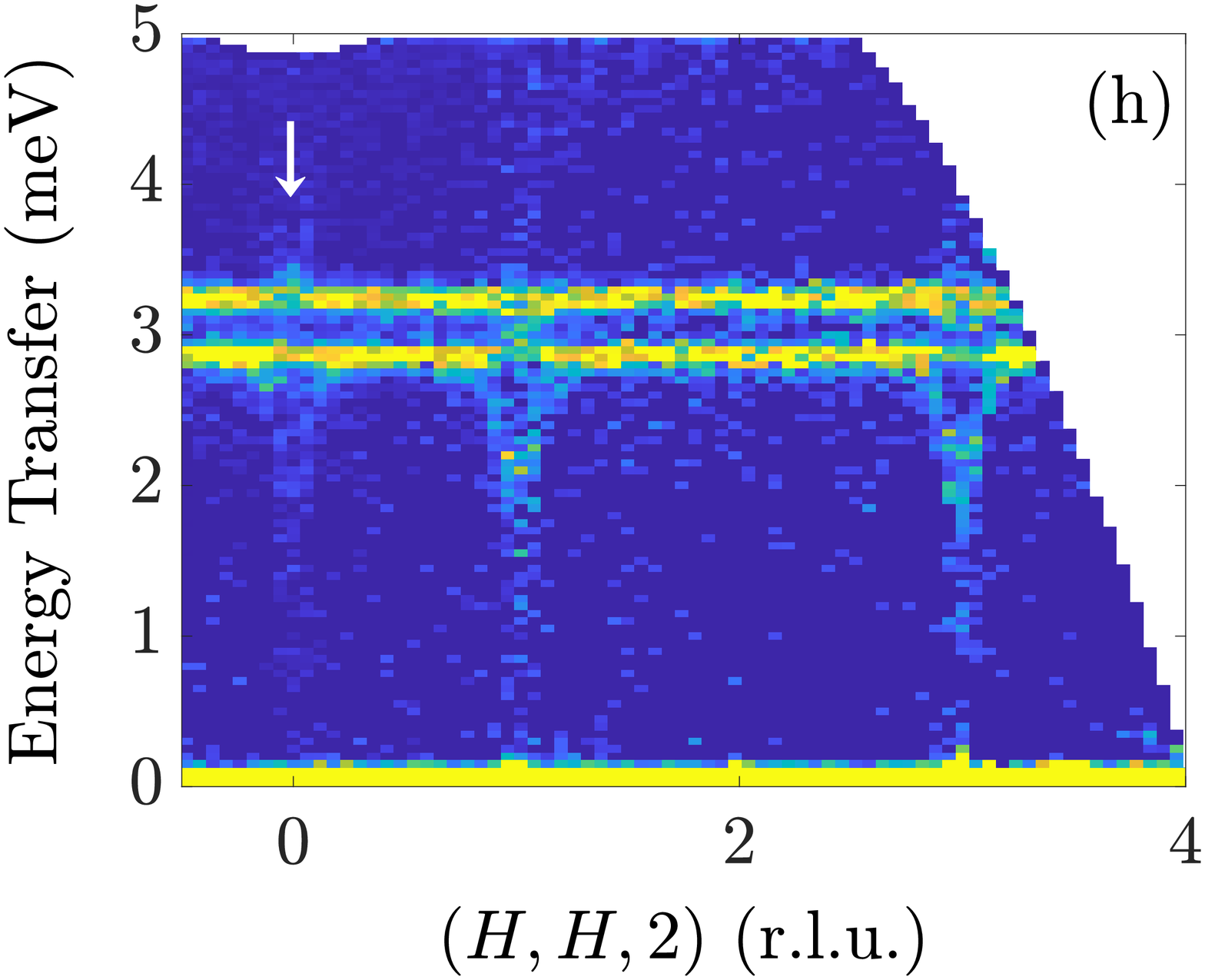}}\hspace{-0.2em}
\subfigure{%
\includegraphics[trim=40 0 0 0, clip,width=5.7cm]{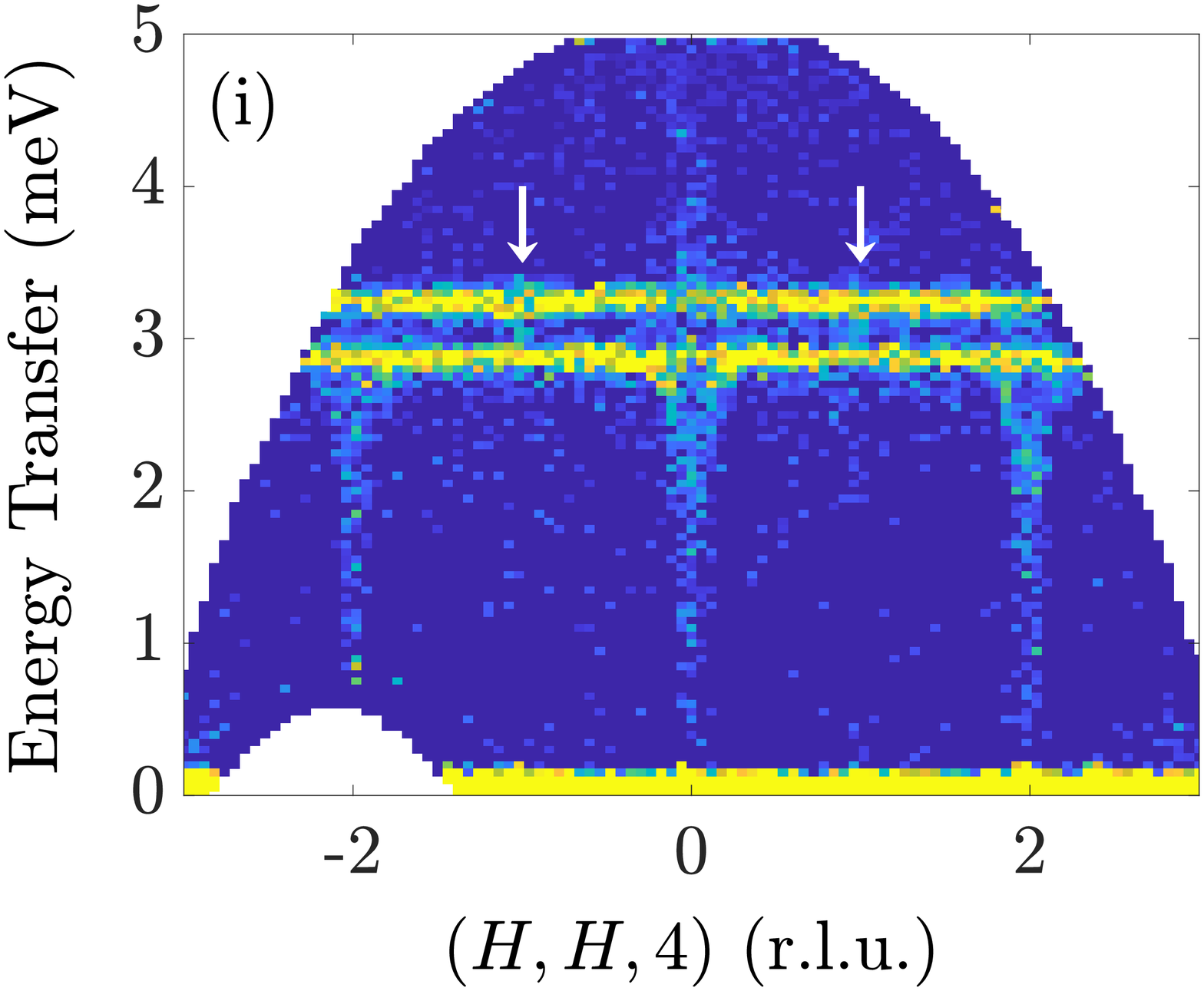}}\vspace{-2.1em}
\subfigure{%
\includegraphics[trim=0 0 30 80, clip,width=6.14cm]{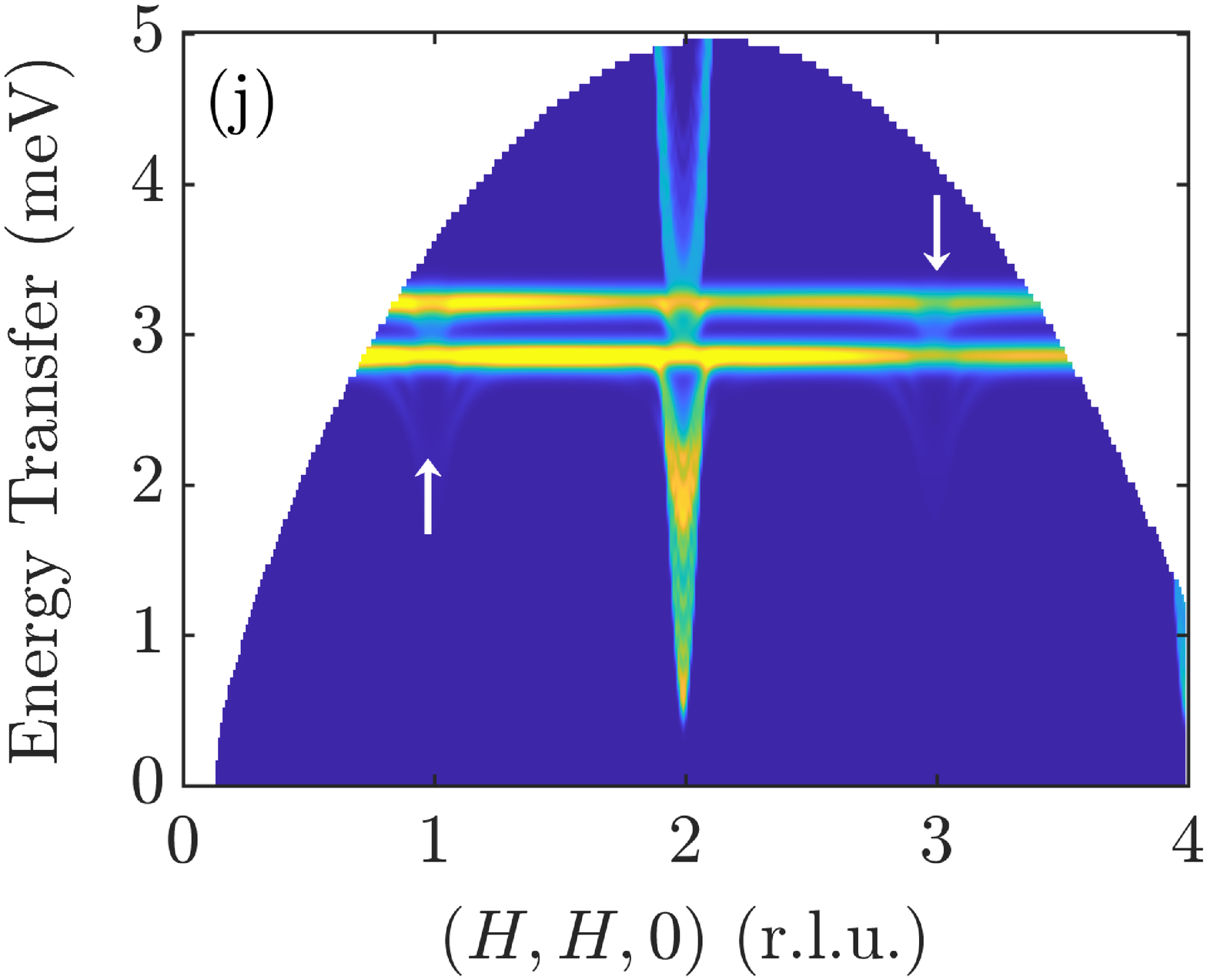}}\hspace{-0.2em} 
\subfigure{%
\includegraphics[trim=50 0 30 80, clip,width=5.7cm]{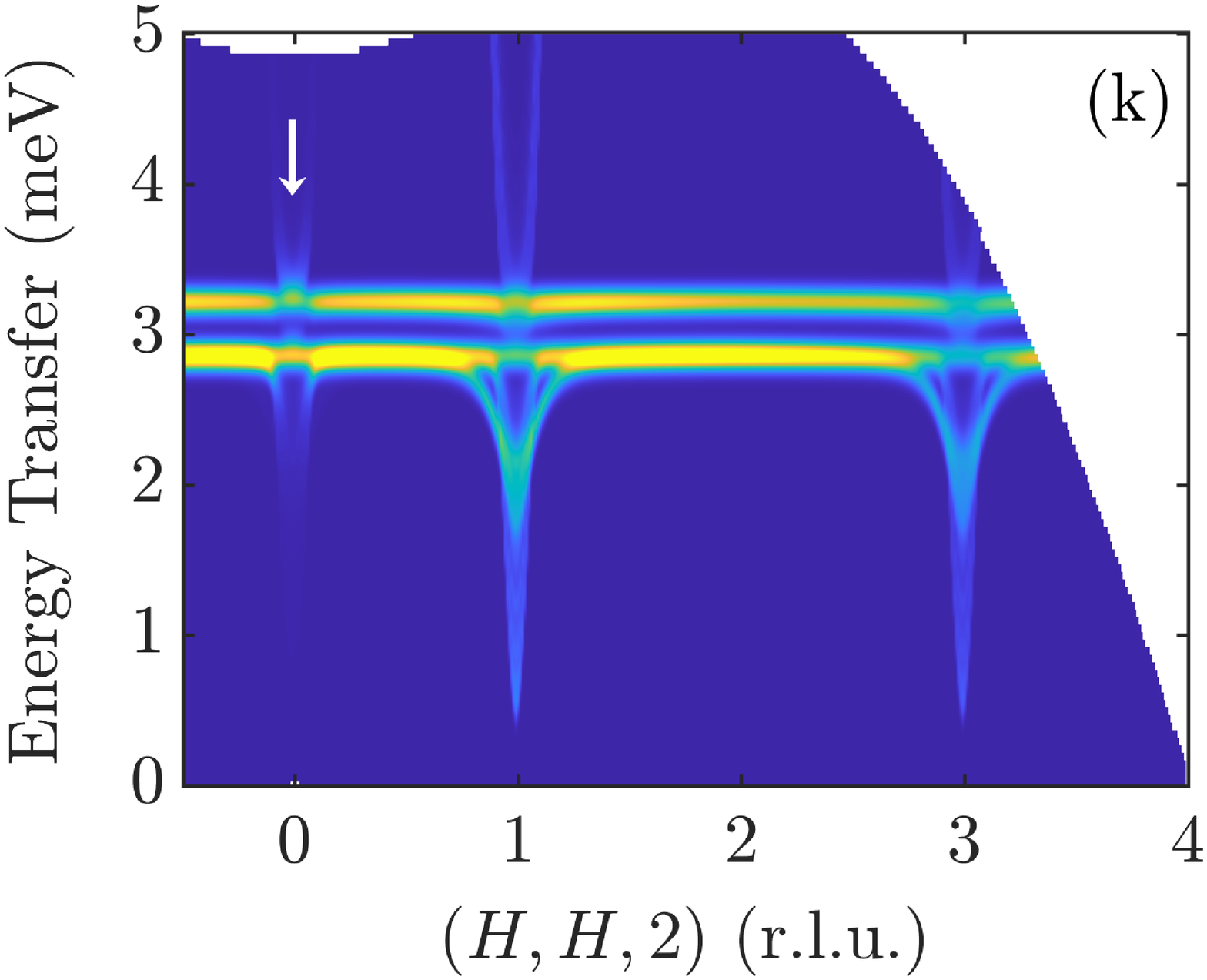}}\hspace{-0.2em} 
\subfigure{%
\includegraphics[trim=50 0 30 80, clip,width=5.7cm]{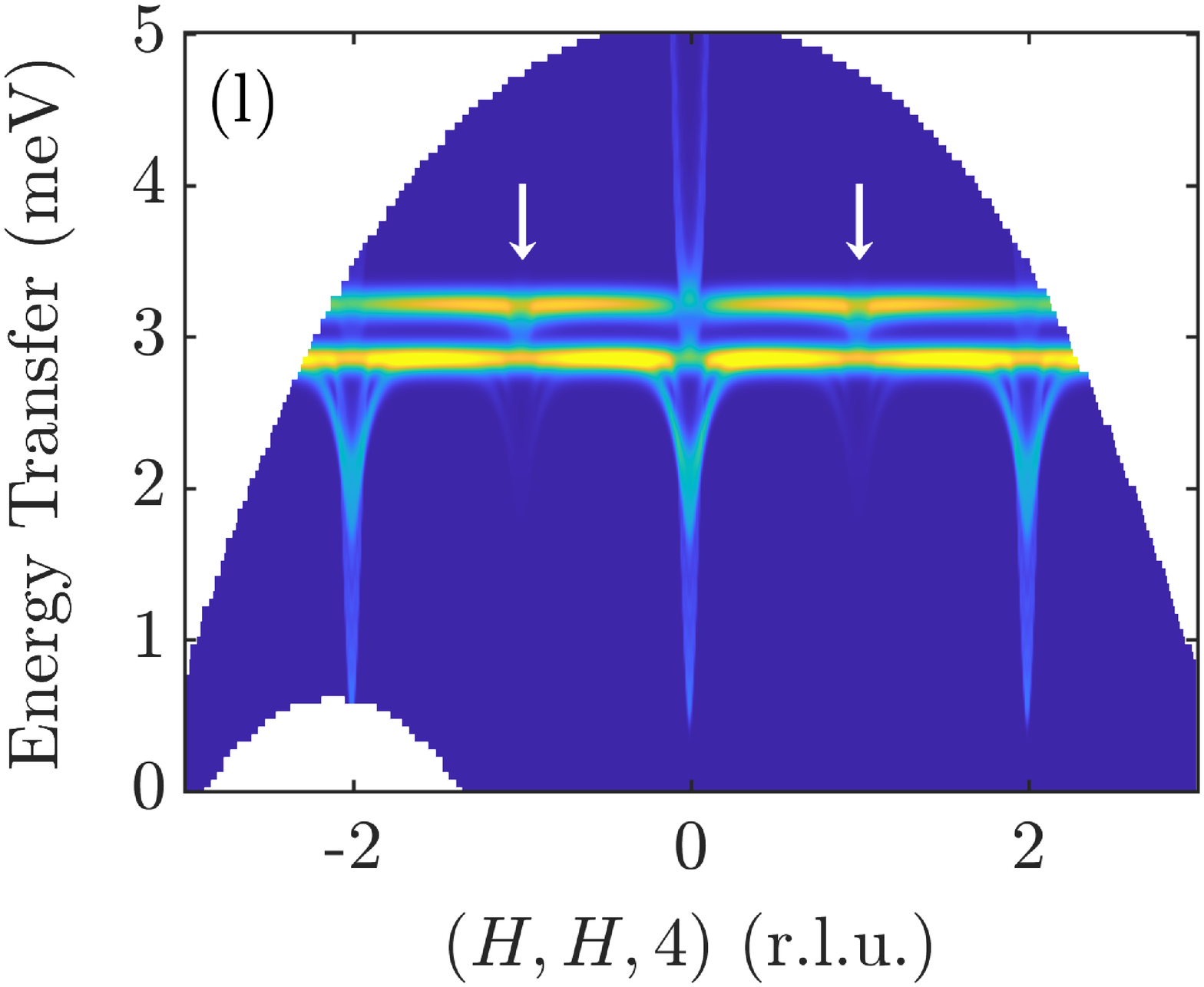}}\vspace{-2.1em} 
\caption{Low-energy inelastic neutron scattering data collected on LET along several directions in reciprocal space, shown intercalated with the calculated spectra for the model Hamiltonian in Eq.~(\ref{H_final}). Panels \hyperref[fig7]{(a)}--\hyperref[fig7]{(c)} and \hyperref[fig7]{(g)}--\hyperref[fig7]{(i)} show experimental data, \hyperref[fig7]{(d)}--\hyperref[fig7]{(f)} and \hyperref[fig7]{(j)}--\hyperref[fig7]{(l)} correspond to calculated spectra using the Fe--Fe exchange model of Shamoto \emph{et al.} \cite{PhysRevB.97.054429}. White arrows denote points of interest in the hybridisation of the modes, reproduced very well by the model.}\label{fig7}
\end{figure*}

In Fig.~\ref{fig7}, spectra simulated from the best-fit model are compared with the LET data for several directions in reciprocal space. Both the experimental and calculated spectra have been averaged over wavevector intervals of $\pm0.1$\,r.l.u. in the directions orthogonal to those shown in the maps (see Appendix \ref{det_LET} for more details). The agreement is seem to be very good. That the model succeeds in describing the hybridisation between the Yb single-ion levels and the Fe spin-wave modes is the central result of this work. Note in particular the hybridisation features highlighted by white arrows, where the Fe spin-wave modes would have zero structure factor in the absence of the Yb--Fe coupling. 

\begin{figure}
\centering
\includegraphics[trim=0 0 0 0, clip,width=8.2cm]{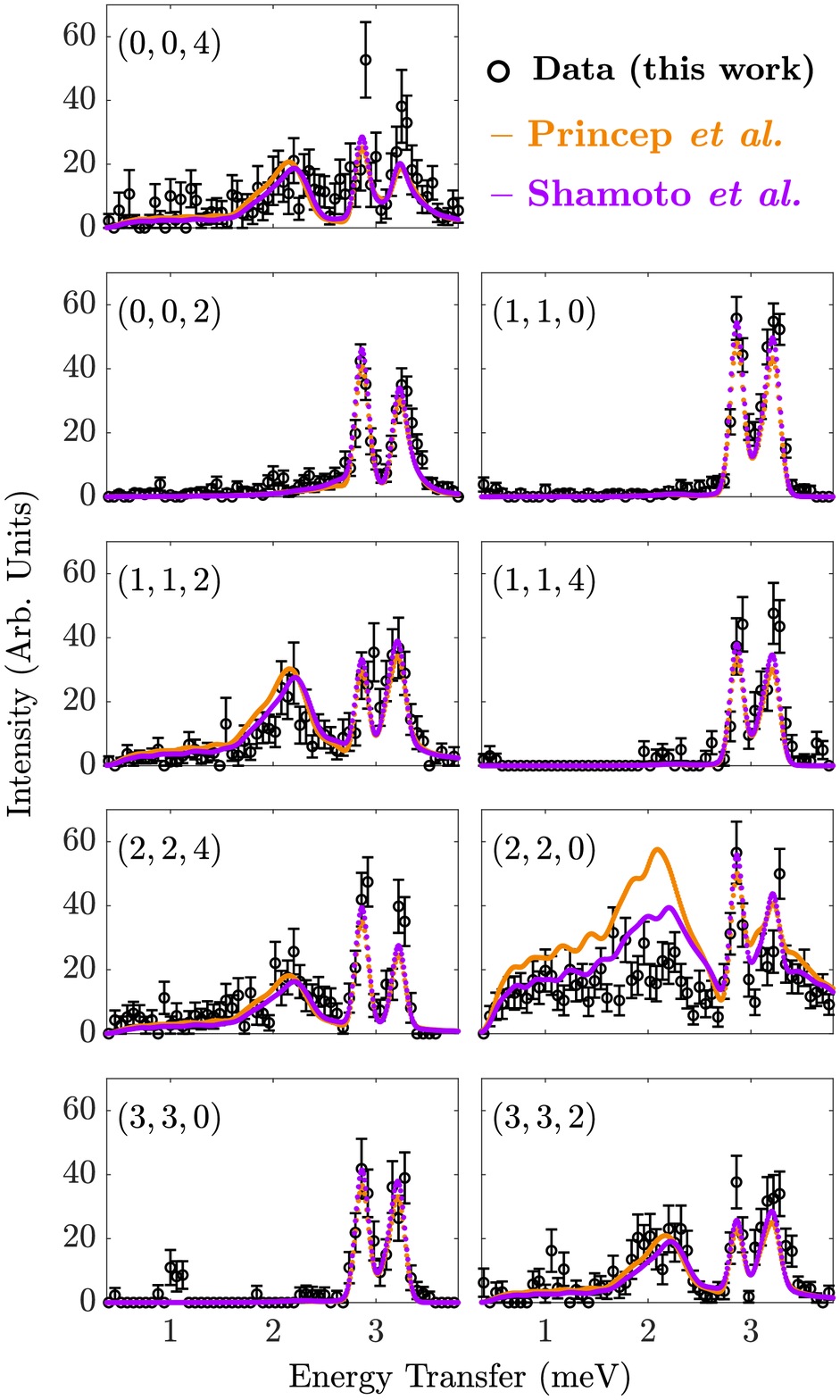}
\caption{Constant-$\mathbf{Q}$ cuts performed on data shown in Fig.~\ref{fig7}. Intensities were averaged over $\pm0.1$ r.l.u. in all directions around the wavevector $\mathbf{Q}$ shown in the panels.}\label{fig8}
\end{figure}

The neutron scattering intensity given by the model is directly compared with the experimental intensities in Fig.~\ref{fig8}. Constant-$\mathbf{Q}$ cuts averaged over a narrow range of wavevector ($\pm0.1$ r.l.u.) are shown along with intensities calculated with our model, integrated over the same wavevector intervals. The centre wavevector of each cut is indicated in the figure labels. The cross-sections were calculated using two sets of Fe--Fe exchange parameters: those reported by Princep \emph{et al.} \cite{Princep} and those of Shamoto \emph{et al.} \cite{PhysRevB.97.054429}. 

\section{Discussion}

\begin{figure}
\centering
\subfigure{%
\hspace{-3em}
\includegraphics[trim=250 40 250 0, clip,width=5.5cm]{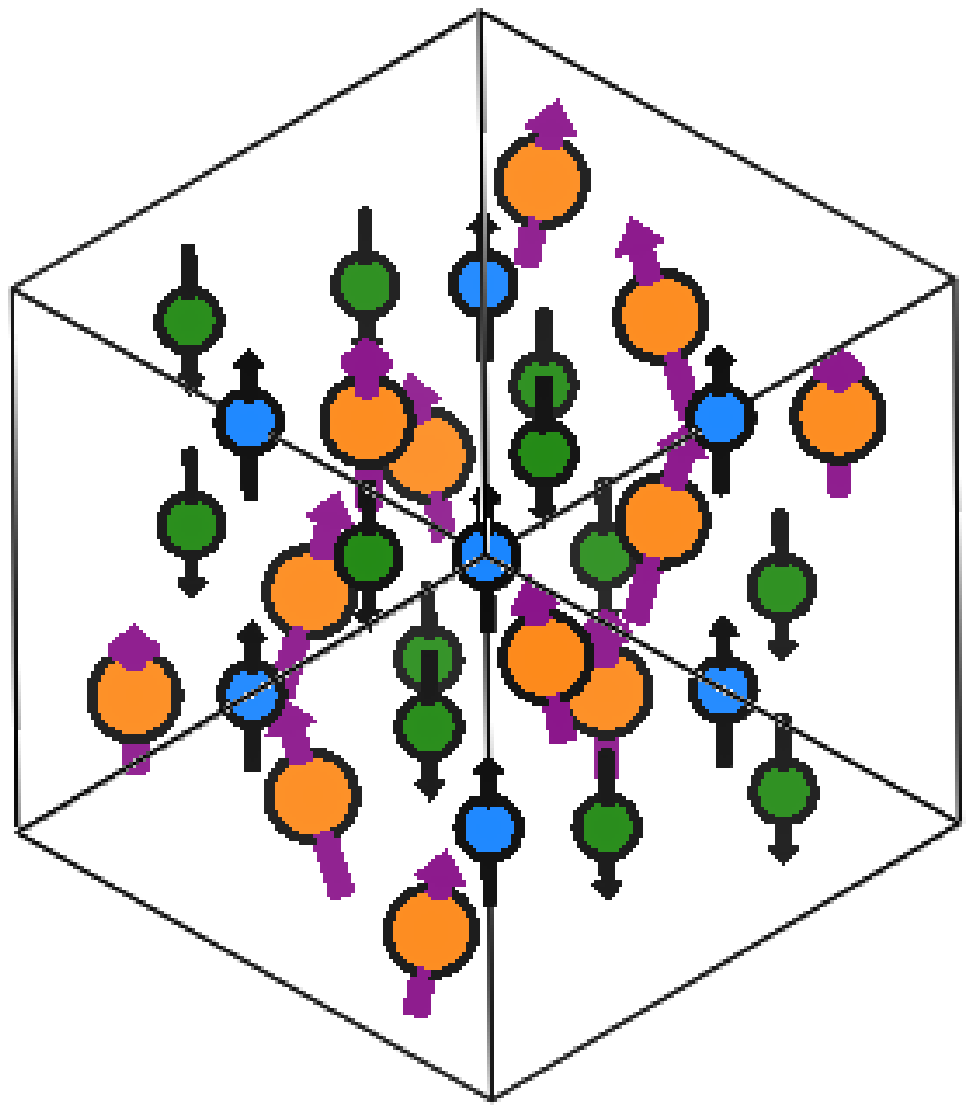}}\hspace{1em} 
\subfigure{%
\includegraphics[trim=350 0 250 0, clip,width=3.5cm]{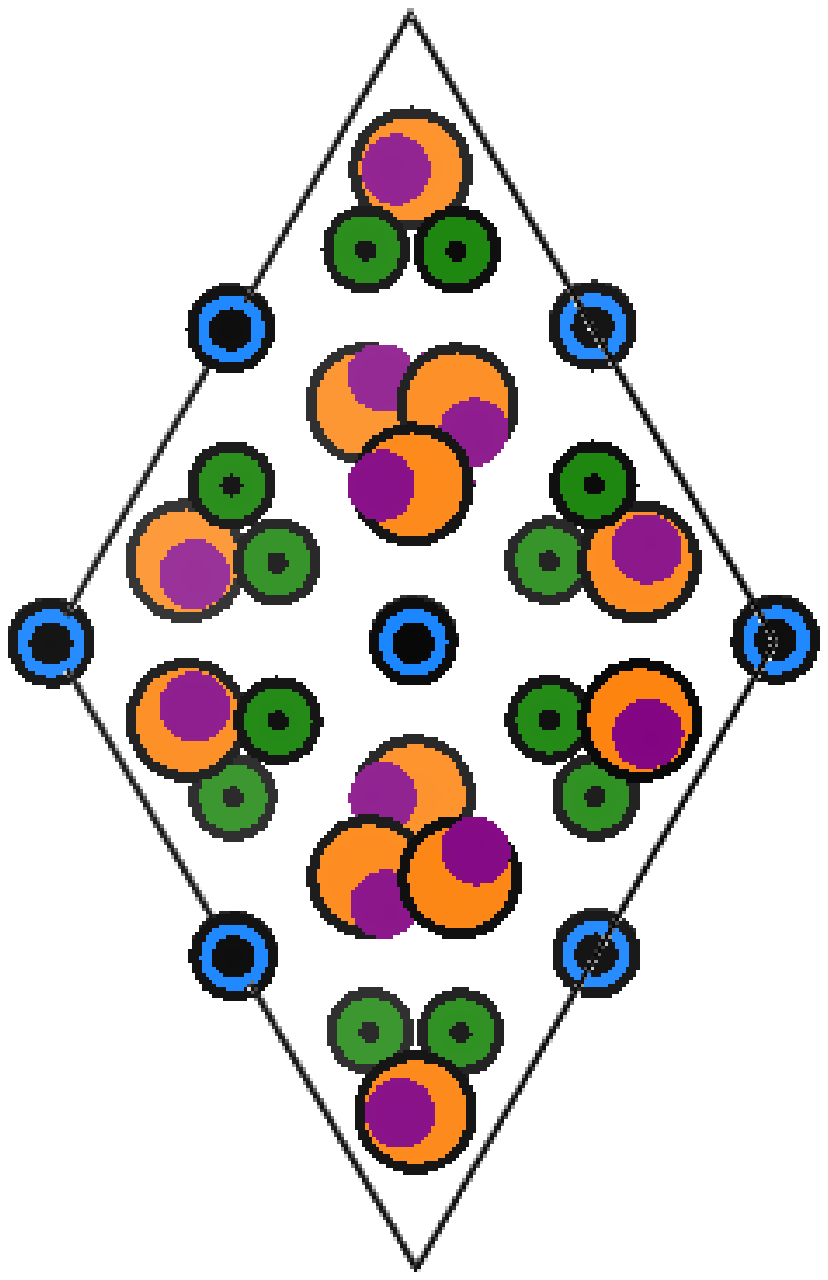}}
\caption{Zero-temperature magnetic structure of Yb$_3$Fe$_5$O$_{12}$, shown down the $[1 1\bar{1}]$ (left) and $[1 1 1]$ (right) axes. The colour code is the same as that of Fig.~\ref{fig1}\subref{fig1-a}. Arrows crossing Yb atoms point along the ground-state magnetic moment of $\mathcal{H}_1$. Size of arrows does not indicate magnetic moment magnitude and is not represented in scale, being used only to denote spin orientation.}\label{fig9}
\end{figure}

The single-ion magnetic moment $\boldsymbol{\mu}_k$ of the Yb atoms may be calculated from the ground state wave-functions of the Hamiltonian $\mathcal{H}_1$. As an example, we state $\boldsymbol{\mu}$ for atoms Yb$_1$ and Yb$_7$, in the local coordinate system
\begin{eqnarray}
\boldsymbol{\mu}_1=(0.61\ \boldsymbol{\xi}_1 - 1.46\ \boldsymbol{\eta}_1)\ \mu_\textup{B}, \nonumber \\
\boldsymbol{\mu}_7=(-0.55\ \boldsymbol{\xi}_7 + 1.86\ \boldsymbol{\zeta}_7)\ \mu_\textup{B} \nonumber.
\end{eqnarray}
Magnetic moments for the other rare-earths in the unit cell can be found by applying the symmetry operators in Appendix~\ref{Yb-local} to $\boldsymbol{\mu}_1$ and $\boldsymbol{\mu}_7$. The single-ion easy axis of the Yb is thus either the local-$\boldsymbol{\eta}$, for atoms of the group 1, or the local-$ \boldsymbol{\zeta}$ for atoms of the group 2. The $T=0$ magnetic structure obtained with our model is depicted in Fig.~\ref{fig9}. The magnetic moment arrangement and magnitude are consistent with the those reported in early neutron scattering studies \cite{Tcheou}. 

The absence of a $\lambda$--shaped feature in the heat capacity data shown in Fig.~\hyperref[fig1]{1(c)} demonstrate that the Yb spin canting is not accompanied by a symmetry breaking. Hence, the magnetic structure presumed to be adopted by the Yb moments in YbIG is not a result of any spontaneous ordering, at least down to about 1.6 K, the lowest temperature at which our measurements were carried out. The canting which we expect to occur in YbIG at low temperatures is due to the competition between the Yb single-ion anisotropy and the exchange interaction with the iron sublattice, which tends to align the Yb magnetic moment along the $\langle 111 \rangle$ directions.

The success of the model in reproducing the intensities of the hybridised excitations, as well as their dispersion, is evident from Fig.~\ref{fig8}. It is noticeable, however, that the agreement is less good for the acoustic mode emerging from the $(2,2,0)$ reciprocal lattice point, where both Shamoto \emph{et al.} and Princep \emph{et al.} models overestimate the spectral weight by a factor of two or more. As stressed earlier, the dispersion and scattering intensity of the iron spin-wave modes observed in our high energy data for YbIG are very similar to those of YIG, and therefore the iron-iron exchange model of Refs.~\cite{Princep,PhysRevB.97.054429} was kept unchanged in our work. The Fe-Fe exchange parameters proposed by Shamoto \emph{et al.} \cite{PhysRevB.97.054429} seem to provide a better agreement with the data in Fig.~\ref{fig7}, especially around $(2,2,0)$. This may suggest that their model is slightly superior in describing the Fe-Fe interactions in the low energy spectrum of YbIG. The model of Princep \emph{et al.}, on the other hand, includes more exchange parameters, and describes better the details of the dispersion of the high energy optical modes in YIG. A more thorough investigation of the Fe--Fe interactions in YbIG is, however, outside the scope of the present work. 

A series of spectroscopic studies performed in the 1960s \cite{PhysRevLett.8.483,PhysRev.122.1376,PhysRevLett.4.123} was successful in measuring the CF splittings of the lowest Kramers doublets of the $J=7/2$ and the $J=5/2$ multiplets of the Yb$^{3+}$ ions in YbIG. The full crystal field spectrum, including all the excited levels of the first $J$ multiplet, was unknown at the time, and later assumed to be identical to that of Yb$_3$Ga$_5$O$_{12}$ \cite{PhysRev.159.251}. In the theoretical studies that followed \cite{PhysRev.135.A155,PhysRev.147.311}, a sophisticated mean-field theory involving the coupling between spin-only (Fe) and spin-orbit (Yb) levels was proposed by Levy. The disadvantage of this approach, made clear in Refs.~\cite{PhysRev.135.A155,PhysRev.147.311}, was the excessive number of unknown model parameters compared to the amount of experimental information available. 

We succeeded in demonstrating that the low temperature magnetic spectrum of Yb$_3$Fe$_5$O$_{12}$ can be very well accounted for with a simpler model for the $4f-3d$ coupling, which is described in terms of an anisotropic exchange field with only three parameters, given in Eq.~(\ref{eq5}). This form of the coupling was first applied to YbIG in Ref.~\onlinecite{PhysRev.122.1376} in order to describe the exchange splitting of the lowest Kramers doublets in the $J=7/2$ and $J=5/2$ levels observed in optical spectroscopic measurements. In our convention, in which the total exchange field is divided by the number of first nearest neighbours $(n=2)$, the values found in Ref.~\onlinecite{PhysRev.122.1376} are $A^{\xi\xi}=0.118\ \textup{meV}$, $A^{\eta\eta}=0.207\ \textup{meV}$ and $A^{\zeta\zeta}=0.233\ \textup{meV}$. These are seen to be very close to the values deduced in our analysis. The model of Ref.~\onlinecite{PhysRev.122.1376} only partially describes the excitations, however, because the details of the coupling, and the effects of the Yb crystal-field excitations upon the low energy magnon spectrum of the Fe sublattices were not considered at the time.

Finally, we believe our work may have ramifications outside the realm of the iron garnets. The rare-earth orthoferrites (REFeO$_3$), as much as the RE iron garnets, have been known and studied for a long time \cite{White}. The cooperative behaviour made possible by the exchange interaction between rare-earth and Fe spins in the REFeO$_3$ compounds is recognised to be of fundamental interest in quantum materials research \cite{Xinwei}. Orthoferrites display two first-order phase transitions, a spin reorientation transition, in addition to a compensation temperature point \cite{White}. Recent inelastic neutron scattering studies on YbFeO$_3$ \cite{PhysRevB.98.064424} and ErFeO$_3$ \cite{Zic} reveal significant hybridisation between RE and Fe modes, similar to those described in this work for Yb$_3$Fe$_5$O$_{12}$. In those works, however, no model for the RE--Fe interaction beyond the mean field was proposed. Despite the complexity of the exchange interactions in orthoferrites, it would be interesting to see if the model developed here could be validated for other compounds presenting strong $d-f$ exchange interactions. 

\section{Conclusion}

Our aim was to develop a model which could describe the inelastic neutron scattering spectrum of Yb$_3$Fe$_5$O$_{12}$.
The very high quality of the neutron data obtained on today's state-of-the-art spectrometers places strong demands on such a model, especially at low energies, where the high resolution of the experimental spectra has brought to light the effects of the $4f-3d$ hybridisation between CF levels and dispersive iron spin waves. Nevertheless, we have shown that the low-energy spectrum can be reproduced very well with a relatively simple model that describes the coupling in terms of an anisotropic exchange between Yb and Fe. Despite the simplifications adopted in our description, the model is much more general: it can be extended to include many CF levels and RE--RE interactions, and also be applied to other rare-earth ions in garnets and other compounds.

\begin{acknowledgments}
We would like to thank Paolo G. Radaelli for fruitful discussions. We would also like to thank the authors of Ref.~\cite{Princep} for making their data available, and S. J. Cassidy for assistance with the MPMS measurements. This work was supported by the U.K. Engineering and Physical Sciences Research Council (Grants Nos. EP/M020517/1, EP/T027991/1 and EP/R042594/1). Experiments at the ISIS Neutron and Muon Source were supported by a beam-time allocation from the Science and Technology Facilities Council (DOI 10.5286/ISIS.E.RB1920689-3 and 10.5286/ISIS.E.RB1910605-3).
\end{acknowledgments}

\appendix

\section{\label{Yb-local}Local coordinate frame of the rare-earth sublattice}
 
The coordinates of the atoms Yb$_k$ inside our choice of primitive unit cell are given in Table~\ref{tab1-supp}. All the coordinates are expressed in the basis of the conventional, orthogonal bcc axes.

\begin{table}
\begin{ruledtabular}
\begin{tabular}{l c c c}
\multicolumn{4}{c}{Yb$_k$} \\
\colrule
$k$ (Group) & x & y & z \\
\colrule
1 (1) & 0 & 3/4 & 7/8 \\
2 (2) & 1/4 & 5/8 & 1\\
3 (1) & 3/8 & 1/2 & 5/4\\
4 (1) & 1/4 & 1/8 & 1\\
5 (2) & 1/2 & 1/4 & 7/8\\
6 (2) & 1/8 & 1/2 & 3/4\\
7 (2) & 0 & 1/4 & 5/8\\
8 (2) & 3/8 & 0 & 3/4\\
9 (1) & 1/2 & -1/4 & 5/8 \\
10 (2) & 1/4 & -1/8 & 1/2 \\
11 (1) & 1/4 & 3/8 & 1/2 \\
12 (1) & 1/8 & 0 & 1/4\\
\end{tabular}
\end{ruledtabular}
\caption{Coordinates of the rare-earth atoms in the primitive unit cell of the garnet lattice, expressed in the basis of the bcc conventional axes.}\label{tab1-supp}
\end{table}
 
\begin{figure}
\centering
\includegraphics[angle=270,trim=100 150 100 150, clip, width=10cm]{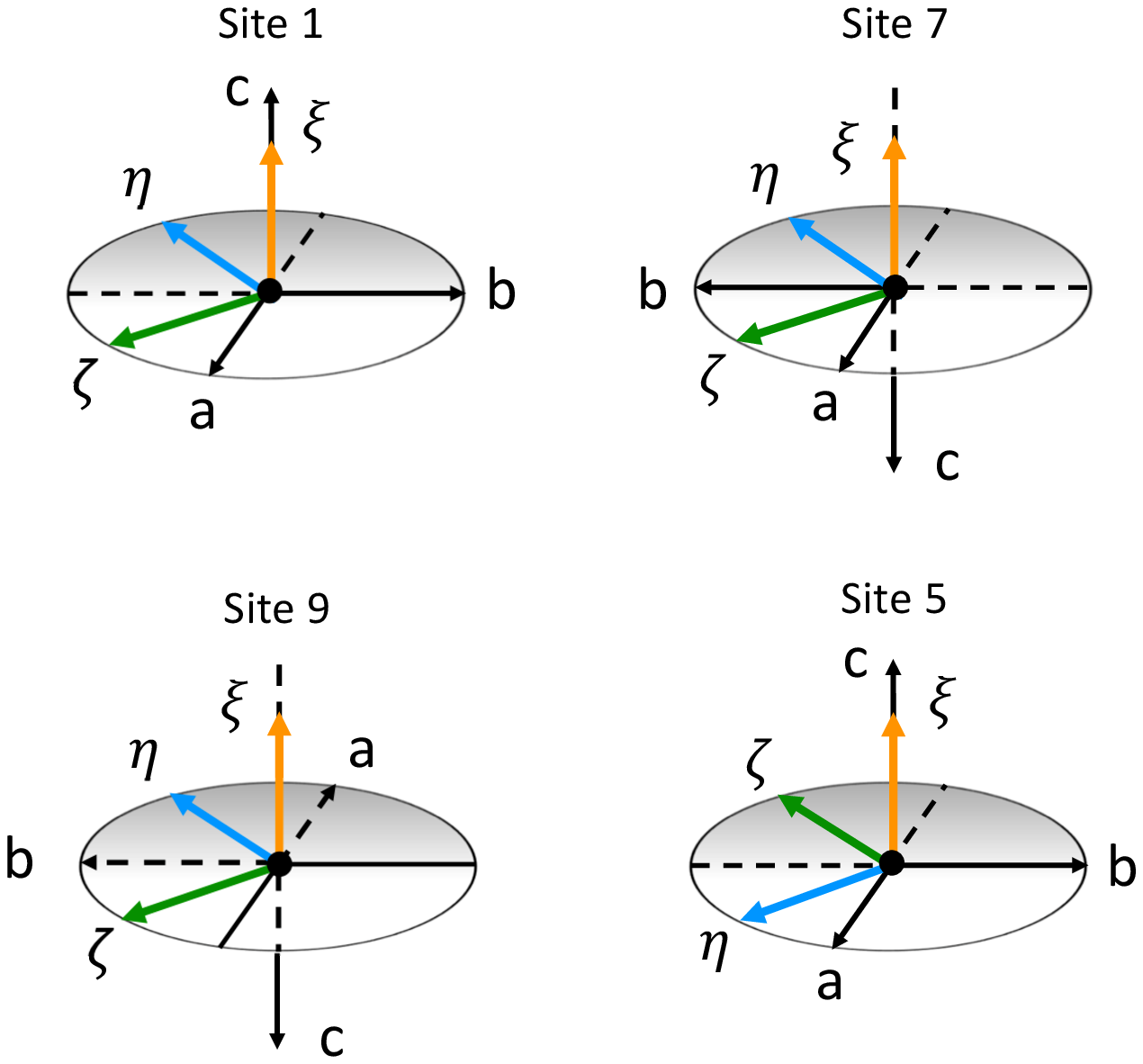}
\caption{Representation of local and global coordinate systems for atoms 1, 5, 7 and 9. Coordinates for the remaining 8 atoms in the primitive unit cell may be obtained by successive 3-fold rotations about the $\left[ 111 \right]$ direction.}\label{fig1-supp}
\end{figure}

The principal axes of the Wybourne operators which describe the crystal field acting on each Yb ion are specified by the local coordinates $\boldsymbol{\xi},\boldsymbol{\eta},\boldsymbol{\zeta}$. The transformation between the local coordinates $\boldsymbol{\xi}_{1},\boldsymbol{\eta}_{1},\boldsymbol{\zeta}_{1}$ of Yb$_{1}$ and the global crystallographic basis $\mathbf{a},\mathbf{b},\mathbf{c}$ is given in Eq.~(\ref{1}). For the atom Yb$_{7}$, 

\begin{align}
\begin{pmatrix}
\boldsymbol{\xi}_7\\
\boldsymbol{\eta}_7 \\
\boldsymbol{\zeta}_7 \\
\end{pmatrix}
=
\begin{pmatrix}
0 \ & \ 0 \ & \ -1 \\
-\frac{1}{\sqrt{2}} \ & \ \frac{1}{\sqrt{2}} \ & \ 0 \\
\frac{1}{\sqrt{2}} \ & \ \frac{1}{\sqrt{2}} \ & \ 0 \\
\end{pmatrix}
\begin{pmatrix}
\mathbf{a}\\
\mathbf{b}\\
\mathbf{c}\\
\end{pmatrix}. \label{2}
\end{align}
Sites for which $k=9,5$ are connected to $k=1, 7$ by an inversion operation 
\begin{align}
I=
\begin{pmatrix}
-1 \ \ & \ \ 0 \ \ & \ \ 0 \\
\ 0 \ \ & \  -1 \ \ & \ \ 0 \\
\ 0 \ \ & \ \ 0 \ \ & \  -1 \\
\end{pmatrix}.
\end{align}
Writing $\mathbf{R_k}=\begin{pmatrix} \boldsymbol{\xi}_k & \boldsymbol{\eta}_k & \boldsymbol{\zeta}_k \end{pmatrix}^{\mathrm{T}}$,
\begin{align}
I \mathbf{R}_1 \rightarrow \mathbf{R}_9& \ \mathrm{and} \ I \mathbf{R}_7 \rightarrow \mathbf{R}_5,
\label{3}
\end{align} 
as represented in Fig.~\ref{fig1-supp}. Local coordinates of all the remaining sites are related to Eqs.~(\ref{1}), (\ref{2}) and (\ref{3}) by $3$-fold rotations about $\left[ 111 \right]$ as follows:
\begin{align*}
C_3 \mathbf{R}_1 \rightarrow   \mathbf{R}_3,& \ \ C_3 \mathbf{R}_3 \rightarrow  \mathbf{R}_{11},  \\
C_3 \mathbf{R}_5 \rightarrow   \mathbf{R}_8 , & \ \  C_3 \mathbf{R}_8 \rightarrow  \mathbf{R}_{10} , \\
C_3 \mathbf{R}_7 \rightarrow   \mathbf{R}_6 , & \ \ C_3 \mathbf{R}_6  \rightarrow   \mathbf{R}_{2} ,  \\
C_3 \mathbf{R}_9  \rightarrow   \mathbf{R}_{12}, &  \ \ C_3 \mathbf{R}_{12} \rightarrow  \mathbf{R}_{4} , \\
\end{align*}
where
\begin{align}
\nonumber
C_3=
\begin{pmatrix}
0 \ \ & \ \ 0 \ \ & \ \ 1 \\
1 \ \ & \ \ 0 \ \ & \ \ 0 \\
0 \ \ & \ \ 1 \ \ & \ \ 0 \\
\end{pmatrix}
.
\end{align}
Note that, in our representation, the $\boldsymbol{\xi}$ direction always lies along the crystallographic $\mathbf{a},\mathbf{b}$ or $\mathbf{c}$.

\section{\label{exchangeFe-Fe}Details about $\mathcal{H}^\textup{Ex}_{ij}$}

In order to calculate the spin-wave spectrum, we need to define a local Cartesian basis for the Fe spins. If ${S}_i^a$(${S}_j^a$), ${S}_i^b$(${S}_j^b$) and ${S}_i^c$(${S}_j^c$) are the spin components of an atom at $\mathbf{r}_i$($\mathbf{r}_j$), projected along the axes of the conventional cubic (global) coordinate system, we can choose for atoms at the $16a$ sites,
\begin{align}
\begin{pmatrix}
{S}_i^x \\
{S}_i^y \\ 
{S}_i^z 
\end{pmatrix}
=
\begin{pmatrix}
-\frac{1}{\sqrt{6}} \ & \ -\frac{1}{\sqrt{6}} \ & \frac{2}{\sqrt{6}} \\
\frac{1}{\sqrt{2}} \ & \ -\frac{1}{\sqrt{2}} \ & \ 0 \\
\frac{1}{\sqrt{3}} \ & \ \frac{1}{\sqrt{3}} \ & \ \frac{1}{\sqrt{3}} 
\end{pmatrix}
\begin{pmatrix}
{S}_i^a \\
{S}_i^b \\ 
{S}_i^c 
\end{pmatrix}. \label{eq3-supp}
\end{align}
The transformation for atoms at $24d$ sites will then be 
\begin{align}
\begin{pmatrix}
{S}_j^x \\
{S}_j^y \\ 
{S}_j^z 
\end{pmatrix}
=
\begin{pmatrix}
\frac{1}{\sqrt{6}} \ & \ \frac{1}{\sqrt{6}} \ & -\frac{2}{\sqrt{6}} \\
\frac{1}{\sqrt{2}} \ & \ -\frac{1}{\sqrt{2}} \ & \ 0 \\
-\frac{1}{\sqrt{3}} \ & \ -\frac{1}{\sqrt{3}} \ & \ -\frac{1}{\sqrt{3}} 
\end{pmatrix}
\begin{pmatrix}
{S}_j^a \\
{S}_j^b \\ 
{S}_j^c 
\end{pmatrix}. \label{eq4-supp}
\end{align}
After performing the rotations given in Eqs.~(\ref{eq3-supp}) and (\ref{eq4-supp}), the Fe spin components are replaced by the standard Holstein-Primakoff bosons. For transition metal ions at $16a$ positions, 
\begin{align}
\renewcommand{\arraystretch}{1.2}
\begin{pmatrix}
{S}_i^x \\
{S}_i^y \\ 
{S}_i^z 
\end{pmatrix}
=
\begin{pmatrix}
\renewcommand{\arraystretch}{1.2}
\frac{\sqrt{2S}}{2}(a_i+a_i^\dagger) \\
-i\frac{\sqrt{2S}}{2}(a_i-a_i^\dagger) \\
S-a_i^\dagger a_i
\end{pmatrix}
, \label{eq5-supp}
\end{align}
while, for atoms at $24d$ positions
\begin{align}
\renewcommand{\arraystretch}{1.2}
\begin{pmatrix}
{S}_j^x \\
{S}_j^y \\ 
{S}_j^z 
\end{pmatrix}
=
\begin{pmatrix}
\renewcommand{\arraystretch}{1.2}
\frac{\sqrt{2S}}{2}(d_j+d_j^\dagger) \\
-i\frac{\sqrt{2S}}{2}(d_j-d_j^\dagger) \\
S-d_j^\dagger d_j
\end{pmatrix}
. \label{eq7-supp}
\end{align}
The Fourier transforms of the bosons in Eq.~(\ref{eq5-supp}) are
\begin{align}
\begin{split}
a_i=\frac{1}{\sqrt{N}}\sum_\mathbf{q} \exp(\textup{i}\mathbf{q}\cdot\mathbf{r}_i) a_i(\mathbf{q}) , \\ a_i^\dagger=\frac{1}{\sqrt{N}}\sum_\mathbf{q} \exp(-\textup{i}\mathbf{q}\cdot\mathbf{r}_i) a_i^\dagger(\mathbf{q}).
\end{split}\label{eq24}
\end{align}
Operators $d_j$ and $d_j^\dagger$ may be similarly defined for the $24d$ site. 

\section{\label{exchangeYb-Fe}The Hamiltonian $\mathcal{H}^\textup{Ex}_{ij}$ expressed in local coordinates}

In matrix form, Eq.~(\ref{HH}) may be written
\begin{align}
 \mathcal{H}^\textup{Ex}_{jk} =
\begin{pmatrix}
{S}_j^{\xi} \ & \ {S}_j^{\eta} \ & \ {S}_j^{\zeta} 
\end{pmatrix}
\begin{pmatrix}
A^{\xi\xi} \ & \ 0 \ & \ 0 \\
0 \ & \ A^{\eta\eta}\ & 0 \\
0 \ & \ 0 \ & \ A^{\zeta\zeta} \\
\end{pmatrix}
\begin{pmatrix}
{J}_k^{\xi} \\
{J}_k^{\eta}\\
{J}_k^{\zeta} \\
\end{pmatrix}, && 
\end{align}

Note that in this representation, the spin operators $\mathbf{S}_j$ of the Fe atoms are written in the local coordinates of the Yb$_k$ to which they couple (see Appendix~\ref{Yb-local}). Before applying the Holstein-Primakoff transformation, ${S}_j^{\xi},{S}_j^{\eta},{S}_j^{\zeta}$ must be transformed to ${S}_j^x,{S}_j^y,{S}_j^z$, the spin components defined in Eqs.~(\ref{eq5-supp}) and (\ref{eq7-supp}). The transformation of the $24d$ Fe spin components between local and global coordinates follows Eq.~(\ref{eq4-supp}). The Yb local coordinates for all the rare-earth atoms in the primitive unit cell were defined in Appendix~\ref{Yb-local}. As an example, we show below how to obtain the rotated exchange matrix for an atom Fe$_j$ bound to Yb$_1$. Using Eqs.~(\ref{1}) and (\ref{eq4-supp}),
\begin{widetext}
\begin{eqnarray}
\mathcal{H}^\textup{Ex}_{j1}
=&&
\begin{pmatrix}
{S}_j^{a} \ & \ {S}_j^{b} \ & \ {S}_j^{c} 
\end{pmatrix}
\begin{pmatrix}
0 \ & \ -\frac{1}{\sqrt{2}} \ & \ \frac{1}{\sqrt{2}} \\
0 \ & \ -\frac{1}{\sqrt{2}} \ & \ -\frac{1}{\sqrt{2}} \\
1\ & \ 0 \ & \ 0 \\
\end{pmatrix}
\begin{pmatrix}
A^{\xi\xi} \ & \ 0 \ & \ 0 \\
0 \ & \ A^{\eta\eta}\ & 0 \\
0 \ & \ 0 \ & \ A^{\zeta\zeta} \\
\end{pmatrix}
\begin{pmatrix}
{J}_1^{\xi} \\
{J}_1^{\eta}\\
{J}_1^{\zeta} \\
\end{pmatrix}
\nonumber \\
=&&
\begin{pmatrix}
{S}_j^{x} \ & \ {S}_j^{y} \ & \ {S}_j^{z} 
\end{pmatrix}
\begin{pmatrix}
\frac{1}{\sqrt{6}} \ & \ \frac{1}{\sqrt{6}} \ & \ -\frac{2}{\sqrt{6}} \\
\frac{1}{\sqrt{2}} \ & \ -\frac{1}{\sqrt{2}} \ & \ 0 \\
-\frac{1}{\sqrt{3}} \ & \ -\frac{1}{\sqrt{3}} \ & \ -\frac{1}{\sqrt{3}}
\end{pmatrix}
\begin{pmatrix}
0 \ & \ -\frac{1}{\sqrt{2}} \ & \ \frac{1}{\sqrt{2}} \\
0 \ & \ -\frac{1}{\sqrt{2}} \ & \ -\frac{1}{\sqrt{2}} \\
1\ & \ 0 \ & \ 0 \\
\end{pmatrix}
\begin{pmatrix}
A^{\xi\xi} \ & \ 0 \ & \ 0 \\
0 \ & \ A^{\eta\eta}\ & 0 \\
0 \ & \ 0 \ & \ A^{\zeta\zeta} \\
\end{pmatrix}
\begin{pmatrix}
{J}_1^{\xi} \\
{J}_1^{\eta}\\
{J}_1^{\zeta} \\
\end{pmatrix}
\nonumber \\
=&&
\begin{pmatrix}
{S}_j^{x} \ & \ {S}_j^{y} \ & \ {S}_j^{z} 
\end{pmatrix}
\widetilde{\mathbf{A}}_{j1}
\begin{pmatrix}
{J}_1^{\xi} \\
{J}_1^{\eta}\\
{J}_1^{\zeta} \\
\end{pmatrix}, \qquad
 \label{eq8-supp}
\end{eqnarray}
\end{widetext}
where
\begin{align}
\widetilde{\mathbf{A}}_{j1}= \frac{1}{3}
\begin{pmatrix}
 -\sqrt{6}A^{\xi\xi}\ & \ -\sqrt{3}A^{\eta\eta} \ & \ 0 \\
0 \ & \ 0\ & \ 3A^{\zeta\zeta} \\
-\sqrt{3}A^{\xi\xi}\ & \ \sqrt{6}A^{\eta\eta} \ & \ 0 \\
\end{pmatrix}. \label{eq9-supp}
\end{align}
Similarly, using Eqs.~(\ref{2}) and (\ref{eq4-supp}),
\begin{align}
\widetilde{\mathbf{A}}_{j7}=\frac{1}{3}
\begin{pmatrix}
 \sqrt{6}A^{\xi\xi}\ & \ 0 \ & \ \sqrt{3}A^{\zeta\zeta}\\
0 \ & \ -3A^{\eta\eta}\ & \ 0 \\
\sqrt{3}A^{\xi\xi}\ & \ 0 \ & \ -\sqrt{6}A^{\zeta\zeta} \\
\end{pmatrix}.
\label{eq10-supp}
\end{align}
Generically, 
\begin{align}
\widetilde{\mathbf{A}}_{jk}=
\begin{pmatrix}
a^{11}_{jk}\ & \ a^{12}_{jk} \ & \ a^{13}_{jk}\\
a^{21}_{jk}\ & \ a^{22}_{jk} \ & \ a^{23}_{jk}\\
a^{31}_{jk}\ & \ a^{32}_{jk} \ & \ a^{33}_{jk}\\
\end{pmatrix}.
\end{align}

After the exchange matrices between all the pairs of interacting atoms in the unit cell are found, following the steps exemplified for Yb$_1$ and Yb$_7$ above, it can be shown that the twelve RE ions in the primitive unit cell may be divided into two symmetry inequivalent groups \cite{PhysRev.124.1401}. From Eqs.~(\ref{eq7-supp}) and (\ref{eq8-supp}), the $3d-4f$ exchange Hamiltonian, Eq.~(\ref{eq6}), describing the coupling between an Yb and an Fe on a $24d$ site (we are neglecting coupling to the $16a$ Fe sites) may be written
\begin{widetext}
\begin{align}
\mathcal{H}^\textup{Ex}_{jk}=&
\begin{pmatrix}
0 \ & \ 0 \ & S
\end{pmatrix}
\widetilde{\mathbf{A}}_{jk}
\begin{pmatrix}
{J}_k^{\xi} \\
{J}_k^{\eta}\\
{J}_k^{\zeta}
\end{pmatrix} 
+
\begin{pmatrix}
\renewcommand{\arraystretch}{1.2}
\frac{\sqrt{2S}}{2}(d_j+d_j^\dagger) &
-i\frac{\sqrt{2S}}{2}(d_j-d_j^\dagger) & 
-d_j^\dagger d_j
\end{pmatrix}
\widetilde{\mathbf{A}}_{jk}
\begin{pmatrix}
{J}_k^{\xi} \\
{J}_k^{\eta}\\
{J}_k^{\zeta}
\end{pmatrix}.\label{eq10}
\end{align}
\end{widetext} 
Equations~(\ref{eq2}) and (\ref{eq3}) in the main text are found by substituting (\ref{eq9-supp}) and (\ref{eq10-supp}), respectively, into the first term on the right-hand side of (\ref{eq10}).

\section{Construction and diagonalisation of the quadratic Hamiltonian\label{exchange_hamiltonian}}

We introduce the fermion operators $C_{km}^\dagger$ and $C_{km}$, which create or annihilate, respectively, an electron on a level $m$ at a site $k$ occupied by an Yb$^{3+}$ ion. From these, two product pseudoboson operators $c^\dagger_{km}=C_{km}^\dagger C_{k0}$ and $c_{km}=C_{k0}^\dagger C_{km}$, for $m\ne 0$, are defined. At sufficiently low temperatures, the ground state occupation is $n_{k0}=C_{k0}^\dagger C_{k0} \sim 1$. Under these conditions, the operators $c^\dagger_{km}$, $c_{km}$ satisfy the Bose commutation relations 
\begin{eqnarray}
\lbrack c_{km},c_{k'm} \rbrack = && 0 \nonumber \\ 
\lbrack c_{km},c_{k'n} ^\dagger \rbrack = && C_{k0}^\dagger \overbrace{C_{km} C_{k'n}^\dagger}^\text{ $\delta_{kk'} \delta_{mn} $} C_{k'0} - C_{k'n}^\dagger C_{k'0}C_{k0}^\dagger C_{km} \nonumber \\
= &&\delta_{kk'}\delta_{mn} \underbrace{C_{k0}^\dagger C_{k0}}_\text{$\sim 1$}-\delta_{kk'} \underbrace{C_{k'n}^\dagger C_{km}}_\text{$\sim 0$} \nonumber \\
\approx && \delta_{kk'}\delta_{mn}, \nonumber
\end{eqnarray}
and can be used to write $\mathcal{H}$ in a quadratic form.

Due to the colossal number of magnetic atoms in the unit cell $(8+12+12=32)$, the matrix form of Eq.~(\ref{H_final}) is built in several blocks. 
First, the pseudoboson operators in Eq.~(\ref{H_final}) are Fourier transformed using Eq.~(\ref{eq24}) and 
\begin{align}
\begin{split}
c_{km}=\frac{1}{\sqrt{N}}\sum_\mathbf{q} \exp(\textup{i}\mathbf{q}\cdot\mathbf{r}_k) c_{km}(\mathbf{q}), \\
c_{km}^\dagger=\frac{1}{\sqrt{N}}\sum_\mathbf{q} \exp(-\textup{i}\mathbf{q}\cdot\mathbf{r}_k) c_{km}^\dagger(\mathbf{q}).
\end{split}\label{eq23}
\end{align}
Next, we write
\begin{equation}
\mathcal{H}(\mathbf{q})=\mathrm{X^\dagger}(\mathbf{q}) \mathrm{H}(\mathbf{q}) \mathrm{X}(\mathbf{q}),
\label{ex_H}
\end{equation}
where 
\begin{widetext}
\begin{equation}
\mathrm{X}^\dagger(\mathbf{q})=
\begin{bmatrix}
a_1^{\dagger}(\mathbf{q}), & \ldots, & d_j^{\dagger}(\mathbf{q}), & \ldots, & c_{km}^{\dagger}(\mathbf{q}), & \ldots, & a_i(\mathbf{-q}), & \ldots, & d_j(-\mathbf{q}), & \ldots & c_{12,1}(-\mathbf{q})
\end{bmatrix}
\end{equation}
\end{widetext}
is the hermitian conjugate of $\mathrm{X}$. As defined in the main text, operators $a_i,d_j$ and $c_{km}$ refer to Fe$^{3+}$ at $16a$, $24d$ and Yb$^{3+}$ at $24c$ sites, respectively. Index $i$ runs from 1 up to 8, $j$ and $k$ from 1 to 12. As only one excited level is included in the model, $m=1$. 

The $\mathrm{H}(\mathbf{q})$ matrix is constructed for the coupling between atoms in one unit cell, and is given by 
\begin{widetext}
\begin{align}
 & \, 
 \begin{matrix}
& \ \ \ \ a_{i'}(\mathbf{q}) & \ \ {\cdots} & \ \ \ d_{j'}(\mathbf{q}) & \ \ {\cdots} & \ \  c_{k'm}(\mathbf{q}) &  \ {\cdots} & \ a_{i'}^{\dagger}(\mathbf{-q}) &  \ {\cdots} & \ \ d_{j'}^{\dagger}(-\mathbf{q}) &\ {\cdots} & \  c_{k'm}^{\dagger}(-\mathbf{q}) \ \
 \end{matrix} 
 \nonumber \\ 
 \textup{H}(\mathbf{q})=
 \begin{matrix}
a_{i}^{\dagger}(\mathbf{q}) \\ \vdots \\ d_{j}^{\dagger}(\mathbf{q}) \\ \vdots \\ c_{km}^{\dagger}(\mathbf{q}) \\ \vdots \\ a_{i}(-\mathbf{q}) \\ \vdots \\ d_{j}(-\mathbf{q}) \\ \vdots \\ c_{km}(-\mathbf{q}) \\
 \end{matrix}
 &
 \begin{bmatrix}
	[D_a(\mathbf{q})]_{8\times8} & [\ 0\ ]_{8\times12} & [\ 0\ ]_{8\times12} & [\ 0\ ]_{8\times8} & [M_\textup{AF}(\mathbf{q})]_{8\times12} & [\ 0\ ]_{8\times12} \\
	\phantom{\vdots} & \phantom{\vdots} & \phantom{\vdots} & \phantom{\vdots} & \phantom{\vdots} & \phantom{\vdots} \\
	 [\ 0\ ]_{12\times8} & [D_d(\mathbf{q})]_{12\times12} & [C(\mathbf{q})]_{12\times12} & [M_\textup{AF}(\mathbf{q})]^\dagger_{12\times8} & [\ 0\ ]_{12\times12} & [F(\mathbf{q})]_{12\times12} \\
	\phantom{\vdots} & \phantom{\vdots} & \phantom{\vdots} & \phantom{\vdots} & \phantom{\vdots} & \phantom{\vdots} \\
	[\ 0\ ]_{12\times8} & [C(\mathbf{q})]^{\dagger}_{12\times12} & [D_c]_{12\times12} & [\ 0\ ]_{12\times8} & [F(-\mathbf{q})]_{12\times12} & [\ 0\ ]_{12\times12} \\
	\phantom{\vdots} & \phantom{\vdots} & \phantom{\vdots} & \phantom{\vdots} & \phantom{\vdots} & \phantom{\vdots} \\
	[\ 0\ ]_{8\times8} & [M_\textup{AF}(\mathbf{q})]_{8\times12} & [\ 0\ ]_{8\times12} & [D_a(\mathbf{q})]_{8\times8} & [\ 0\ ]_{8\times12} & [\ 0\ ]_{8\times12} \\
	\phantom{\vdots} & \phantom{\vdots} & \phantom{\vdots} & \phantom{\vdots} & \phantom{\vdots} & \phantom{\vdots} \\
	 [M_\textup{AF}(\mathbf{q})]^\dagger_{12\times8} & [\ 0\ ]_{12\times12} & [F(-\mathbf{q})]_{12\times12}^{\dagger} & [\ 0\ ]_{12\times8} & [D_d(\mathbf{q})]_{12\times12} & [C(-\mathbf{q})]^\dagger_{12\times12} \\
	\phantom{\vdots} & \phantom{\vdots} & \phantom{\vdots} & \phantom{\vdots} & \phantom{\vdots} & \phantom{\vdots} \\
	[\ 0\ ]_{12\times8} & [F(\mathbf{q})]^{\dagger}_{12\times12} & [\ 0\ ]_{12\times12} & [\ 0\ ]_{12\times8} & [C(-\mathbf{q})]_{12\times12} & [D_c]_{12\times12} \\
 \end{bmatrix},
 \label{Hq_mat}
\end{align}
\end{widetext}
where the rows and columns are labelled with the components of $\mathrm{X}^\dagger(\mathbf{q})$ and $\mathrm{X}(\mathbf{q})$. The matrix elements on each block of (\ref{Hq_mat}) are 
\begin{widetext}
\begin{equation}
\begin{split}
[D_a(\mathbf{q})]_{ii'}&= S \left[ -2\mathcal{J}_3 -6\mathcal{J}_6 +6 (\mathcal{J}_1+\mathcal{J}_4) \right]  \delta_{ii'}  +  S  \sum_{\textbf{r}_{i'}\ne \textbf{r}_{i}} \mathcal{J}_{ii'} \exp[-\textup{i}\mathbf{q}\cdot(\mathbf{r}_i-\mathbf{r}_{i'})] . \\[8pt]
[M_\textup{AF}(\mathbf{q})]_{ij'}&=S \mathcal{J}_{ij'} \exp[-\textup{i}\mathbf{q}\cdot(\mathbf{r}_i-\mathbf{r}_{j'})]. \\[8pt]
[D_d(\mathbf{q})]_{jj'}&= \left\{ S \left[ -4\mathcal{J}_2 -8\mathcal{J}_5 +4 (\mathcal{J}_1+\mathcal{J}_4)  \right] - \sum_{k'}K_{jk'}^3 \right\} \delta_{jj'}  + S \sum_{\textbf{r}_{j'}\ne \textbf{r}_j}  \mathcal{J}_{jj'} \exp[-\textup{i}\mathbf{q}\cdot(\mathbf{r}_j-\mathbf{r}_{j'})] , \\
&\mathrm{where} \ K_{jk'}^3=\left( a^{31}_{jk'} {J}^\eta_{00} +a^{32}_{jk'} {J}^\xi_{00}+a^{13}_{jk'} {J}^\zeta_{00} \right).\\[8pt]
[D_c]_{kk'}&=E_{km} \delta_{kk'}  . \\[8pt]
[C(\mathbf{q})]_{jk'}&=\frac{\sqrt{2S}}{2}  K^{2}_{jk'} \exp[-\textup{i}\mathbf{q}\cdot(\mathbf{r}_j-\mathbf{r}_{k'})], \\
&\mathrm{where} \ K^{2}_{jk'} = {J}^\xi_{m0} (a^{11}_{jk'} + \textup{i} a^{21}_{jk'}) + {J}^\eta_{m0} (a^{12}_{jk'} + \textup{i} a^{22}_{jk'}) +{J}^\zeta_{m0} (a^{13}_{jk'} + \textup{i} a^{23}_{jk'}). \\[8pt]
[F(\mathbf{q})]_{jk'}&=\frac{\sqrt{2S}}{2}  K^{1}_{jk'} \exp[-\textup{i}\mathbf{q}\cdot(\mathbf{r}_j-\mathbf{r}_{k'})], \\
&\mathrm{where} \ K^{1}_{jk'} = {J}^\xi_{0m} (a^{11}_{jk'} + \textup{i} a^{21}_{jk'})+{J}^\eta_{0m} (a^{12}_{jk'} + \textup{i} a^{22}_{jk'})+{J}^\zeta_{0m} (a^{13}_{jk'} + \textup{i} a^{23}_{jk'}). \\
\end{split}\label{matrix_elements}
\end{equation}
\end{widetext}
The Bogoliubov transformation $\mathcal{T}(\mathbf{q})$ which diagonalises $\mathrm{H}(\mathbf{q})$
\begin{equation}
\mathcal{D}(\mathbf{q})=\mathcal{T}^\dagger(\mathbf{q}) \mathrm{H}(\mathbf{q}) \mathcal{T}(\mathbf{q})
\end{equation}
is found numerically using the procedure outlined by Colpa in Ref.~\onlinecite{colpa}. 

\section{Neutron scattering cross-section\label{Xsec}}

The cross-section for magnetic neutron scattering is proportional to the dynamical structure factor $S(\mathbf{Q},\omega)$, which may be written \cite{Andrews_book}
\begin{equation}
S(\mathbf{Q},\omega) = \sum_{\lambda_{\textrm{i}}}p_{\lambda_{\textrm{i}}}\sum_{\lambda_{\textrm{f}}}|\langle \lambda_\textrm{f}\left|\textbf{M}_\perp(\textbf{Q})|\lambda_\textrm{i} \rangle \right|^2\delta(E_{\lambda_\textrm{f}}-E_{\lambda_\textrm{i}}-\hbar \omega),
\label{SQw1}
\end{equation}
where $\textbf{M}_\perp(\textbf{Q}) = \hat{\textbf{Q}}\times \textbf{M}(\textbf{Q}) \times \hat{\textbf{Q}}$ is the component of $\textbf{M}(\textbf{Q})$ perpendicular to $\textbf{Q}$, $\textbf{M}(\textbf{Q})$ being the Fourier transform of the magnetization operator, and $\hat{\textbf{Q}}$ the unit vector in the direction of $\textbf{Q}$. Also in Eq.~(\ref{SQw1}), $\lambda_{\textrm{i}}$ and $\lambda_{\textrm{f}}$ are the initial and final states of the system, and $p_{\lambda_{\textrm{i}}}$ is the thermal occupation probability of the initial state.

In the determination of the crystal-field model for the Yb$^{3+}$ single-ion excitations, we assume the dipole approximation for $\textbf{M}(\textbf{Q})$,
\begin{equation}
\textbf{M}(\textbf{Q}) = -\mu_\textrm{B}f_\textrm{Yb}(Q)\exp(-W)(\textbf{L}+2\textbf{S}),
\label{SQw2}
\end{equation}
where $f_\textrm{Yb}(Q)$ is the magnetic form factor of Yb$^{3+}$ and $\exp(-W)$ is the Debye--Waller factor. Note that higher order multipoles in the scattering operator $\textbf{M}(\textbf{Q})$ would need to be included to obtain accurate intensities for the $J=7/2$ to $5/2$ inter-level transitions. Substituting (\ref{SQw2}) into (\ref{SQw1}) we obtain
\begin{eqnarray}
S(\mathbf{Q}, \omega)& = & \mu_\textrm{B}^2 f_\textrm{Yb}^2(Q) \sum_{n} p_{n} \nonumber \sum_{m} |\langle\Gamma_{m} | \mathbf{L}_\perp +2\mathbf{S}_\perp| \Gamma_{n} \rangle |^2 \\
& &\times \delta (E_m-E_n-\hbar \omega),
\label{eq4}
\end{eqnarray}
where $p_n=\exp(-E_n/k_\textup{B} T)$ is the thermal population of level $\Gamma_n$ at a temperature $T$, and $E_n$ and $E_m$ are the energies of the initial and final CF levels, respectively. Because our measurement was performed at low temperature we have put $\exp(-W) = 1$.

For the analysis of the propagating excitations on the coupled Yb and Fe sublattices we express the dipole approximation for the scattering operator as 
\begin{align}
{\textbf{M}(\textbf{Q})}= &\mu_\textrm{B}g_Sf_\textrm{Fe}(Q)\sum_{\alpha=i,j} \textbf{S}_{\alpha}(\textbf{Q}) \nonumber \\
&+\mu_\textrm{B}g_J f_\textrm{Yb}(Q)\sum_k \textbf{J}_k(\textbf{Q}),
\label{SQw3}
\end{align}
where $g_S = 2$, $g_J = 8/7$, and $f_\textrm{Fe}(Q)$ is the magnetic form factor of the Fe$^{3+}$ ions.
The operators $\textbf{S}_j(\textbf{Q})$ and $\textbf{J}_k(\textbf{Q})$ are written in terms of the pseudoboson operators $a_i(\mathbf{q})$, $d_j(\mathbf{q})$, $c_{km}(\mathbf{q})$ and their hermitian conjugates, which are obtained from Eq.~(\ref{eq22}) and the Fourier transforms of (\ref{eq24}) and (\ref{eq23}). The eigenstates $|\lambda\rangle$ of the diagonal Hamiltonian $\mathcal{D}(\mathbf{q})$ are given by $\mathcal{T}(\mathbf{q})\mathrm{X}(\mathbf{q})$. 

\section{\label{det_MAPS}Details on the analysis of the MAPS data} 

\begin{figure}
\centering
\includegraphics[trim=0 0 0 0, clip,width=8.5cm]{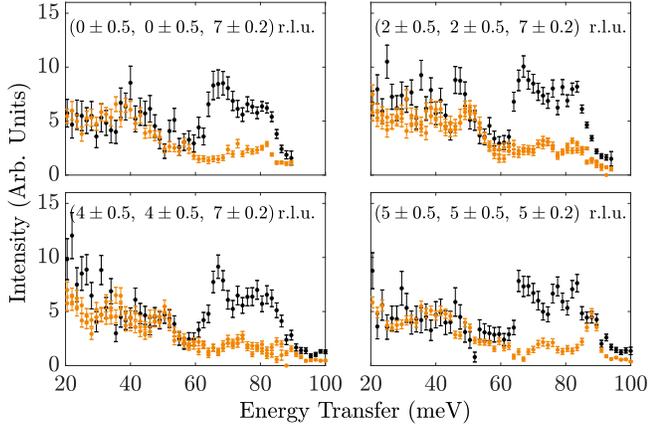}\vspace{-0.7em}
\caption{Procedure used to separate scattering from crystal-field excitations from scattering due to optical Fe spin-wave modes and phonons. Data on Y$_3$Fe$_5$O$_{12}$ (orange) were scaled to data on Yb$_3$Fe$_5$O$_{12}$ (black), and subsequently subtracted from them. The resulting pattern, noticeably within the $[60,90]$\,meV energy interval, is attributed to the single-ion excitations of the rare-earth. Integration ranges are indicated at the top of each panel.}\label{fig2-supp}
\end{figure}

Using the measured spectrum of Y$_3$Fe$_5$O$_{12}$ as a background, we were able to separate rare-earth single-ion excitations from the high energy spin-waves, mostly optical modes, occurring at a similar energy range. To demonstrate that this procedure is indeed reliable, Fig.~\ref{fig2-supp} displays some examples of constant-$\mathbf{Q}$ cuts along different orientations, showing that, after scaling the spin-wave intensities, data on Y$_3$Fe$_5$O$_{12}$ offers a good estimate of the spin-wave and phonon backgrounds on Yb$_3$Fe$_5$O$_{12}$. 

\begin{figure}
\centering
\includegraphics[width=8.5cm]{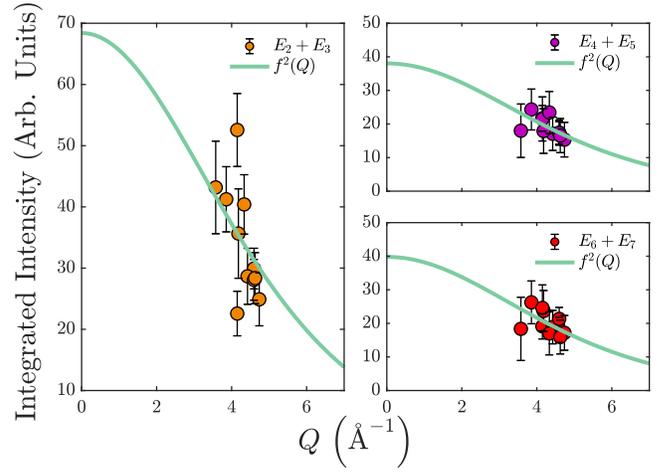}\vspace{-0.7em}
\caption{Integrated intensities, obtained as detailed in the text, of the constant-$\mathbf{Q}$ cuts performed on MAPS data.}\label{fig3-supp}
\end{figure}

The background-subtracted cuts were then fitted assuming three Gaussian components with widths corresponding roughly to the instrumental resolution. Integrated intensities, shown in Fig.~\ref{fig3-supp}, for the levels $(E_2+E_3)$, $(E_4+E_5)$ and $(E_6+E_7)$ are then obtained in arbitrary units. A function $a_if_\mathrm{Yb}^2(Q)$, where $a_i$ is a scaling constant and $i=1,2,3$, is plotted along with the data in Fig.~\ref{fig3-supp}. The normalised integrated intensities at $Q=0$, and energies for each one of the modes are summarised in Table~\ref{tab1} in the main text. 

\section{\label{det_LET}Details on the analysis of the LET data}

The Fe acoustic spin-wave modes are seen in Fig.~\ref{fig7} to be very steep in the energy range of interest to the analysis of the coupling between Yb and Fe, i.e, up to 5\,meV. In other words, the spin-wave dispersion varies considerably for relatively small deviations in wavevector away from the reciprocal lattice points. This needs to
be taken into account when comparing the experimental
data with the simulated spectrum. 

The measured intensity maps shown in Figs.~\hyperref[fig7]{8(a)}--\hyperref[fig7]{8(c)} and \hyperref[fig7]{8(g)}--\hyperref[fig7]{8(i)} are plotted as a function of energy and one wavevector direction $\mathbf{Q}$, and have been averaged over intervals $\Delta \mathbf{Q}_{\perp1}$ and $\Delta \mathbf{Q}_{\perp2}$ in the two directions perpendicular to $\mathbf{Q}$. An identical averaging was applied to the simulated spectrum by dividing the plane defined by $\Delta \mathbf{Q}_{\perp1}$ and $\Delta \mathbf{Q}_{\perp2}$ into bins of area $10^{-4}$\,(r.l.u.)$^2$ and calculating the intensities in each bin. A similar procedure, but with a volume average, was used for the constant-\textbf{Q} cuts displayed in Fig.~\ref{fig8}.

\bibliography{PhysRevApplied.4.047001,
Chumak,
Nakamura,
Geprags,
Hur,
Yoshimoto,
Pearson,
Cherepanov,
Princep,
PhysRevB.97.054429,
Uchida1,
Uchida2,
PhysRevX.4.041023,
PhysRevB.92.224415,
PhysRevLett.117.217201,
Geller1957,
PhysRev.137.A1034,
Guillot1984,
Hock1990,
Hock1991,
Tcheou,
PhysRev.124.1401,
MAPS,
LET,
Ghanathe,
Guillot,
PhysRevLett.4.123,
PhysRev.122.1376,
PhysRev.140.A1944,
Buyers_1971,
PhysRev.167.510,
PhysRev.159.251,
VanVleck,
Hutchings,
PhysRevLett.8.483,
PhysRev.135.A155,
PhysRev.147.311,
White,
PhysRevB.98.064424,
Zic,
colpa,
PhysRevB.89.024409,
PhysRevB.92.054436,
PhysRev.118.1490,
Andrews_book,
Xinwei
}

\end{document}